\RequirePackage{fix-cm}
\documentclass[smallextended]{svjour3}       % onecolumn (second format)
\smartqed  % flush right qed marks, e.g. at end of proof
\usepackage{graphicx}
\usepackage{amsmath,amssymb}
\usepackage{amsfonts}
\usepackage{xspace} 
\usepackage[usenames]{color}
\usepackage{dcolumn}
\usepackage{bm}
\usepackage{mathrsfs}
\usepackage[colorlinks=true]{hyperref}
\usepackage[all]{hypcap} 
\usepackage[utf8]{inputenc}
\usepackage{multirow}
\usepackage{aas_macros} 
\usepackage{hyperref}
\usepackage{cite} % To group references
\newcommand{\eq}{\begin{equation}}
\newcommand{\be}{\begin{equation}}
\newcommand{\eeq}{\end{equation}}
\newcommand{\ee}{\end{equation}}
\newcommand\ba{\begin{eqnarray}}
\newcommand\ea{\end{eqnarray}}

\newcommand{\nn}{\nonumber}

\begin{document}

\title{Neutron star tidal deformability and equation of state constraints
}

%\titlerunning{Neutron Star Tidal Deformability and Equation of State Constraints}        

\author{Katerina Chatziioannou}


\institute{K. Chatziioannou \at
Center for Computational Astrophysics, Flatiron Institute, 162 5th Ave, New York, NY 10010
}

\date{Received: date / Accepted: date}

\maketitle

\begin{abstract}
Despite their long history and astrophysical importance, some of the key properties of neutron stars are still uncertain. 
The extreme conditions encountered in their interiors, involving matter of uncertain composition at extreme 
density and isospin asymmetry, 
uniquely determine the stars' macroscopic properties within General Relativity. 
Astrophysical constraints on those macroscopic properties, 
such as neutron star masses and radii, 
have long been used to understand the microscopic properties of the matter that forms them. 
In this article we discuss another astrophysically observable macroscopic property of neutron stars that can be used to study their 
interiors: their tidal deformation. Neutron stars, much like any other extended object with structure, are tidally deformed
when under the influence of an external tidal field. In the context of coalescences of neutron stars observed through
their gravitational wave emission, this deformation, quantified through a parameter termed the \emph{tidal deformability},
can be measured. We discuss the role of the tidal deformability in observations of coalescing neutron stars with gravitational
waves and how it can be used to probe the internal structure of Nature's most compact matter objects. 
Perhaps inevitably, a large portion of the discussion will be dictated by GW170817, the most informative 
confirmed detection of a binary neutron star coalescence
with gravitational waves as of the time of writing.
\end{abstract}

\tableofcontents

%%%%%%%%%%%%%%%%%%%%%%%%%%%%%%%
\section{Introduction}
\label{intro}

Neutron stars are fascinating astrophysical objects almost as much as they are complicated. Their interiors host matter in some of the most extreme 
conditions imaginable, with densities and neutron-proton asymmetry exceeding those typically encountered in the nuclei of atoms. The unique combination of supranuclear densities and low 
temperatures, unattainable anywhere else in the Universe, makes neutron stars coveted and unique 
laboratories for testing extreme matter.
The properties of neutron star interiors, usually expressed in terms of a relation between the interior pressure $p$, the energy density $\epsilon$, and the temperature $T$
is the subject of research for nuclear physicists, astrophysicists, gravitational wave physicists, and data analysts alike~\cite{Lattimer:2015nhk,Ozel:2016oaf,Watts:2016uzu,Paschalidis:2016vmz,Oertel:2016bki,Miller:2016pom,Baym:2017whm,Degenaar:2018lle}.

 This relation, $p(\epsilon,T)$, commonly referred to as the \emph{equation of state} of neutron stars, 
 involves their microscopic properties
 and uniquely determines their macroscopic 
 properties, such as the maximum neutron star mass possible, and the stars' sizes and tidal properties~\cite{1992ApJ...398..569L}. 
 Over the years, a large number of models for the equation of state have been calculated and
 proposed by the nuclear physics community. Models differ by the neutron star composition they entail, the nucleon interaction
 properties, as well as the methodologies involved to tackle the corresponding many-body Schroedinger equation.
 Perhaps unsurprisingly, the outcome is a wide range of possible astrophysical properties for neutron stars, culminating in, 
 for example, an almost 50\% variation on their predicted size~\cite{Ozel:2016oaf}, see also Fig.~\ref{fig:EoSs}. 
 This sizeable variation introduces 
 non negligible uncertainties 
 in numerous astrophysical and relativistic phenomena that involve neutron stars, such as supernova, neutron star coalescences, gamma ray bursts, the production
 of heavy elements, and tests of strong-field General Relativity, to name but a few.
 
The large interdisciplinary interest in the microscopic and macroscopic properties of neutron stars has led to considerable effort to constrain them observationally and experimentally.
This involves lower limits on the maximum neutron star mass obtained by the detection of heavy pulsars in radio surveys~\cite{Demorest:2010bx,Antoniadis:2013pzd,Cromartie:2019kug};
constraints on radii using electromagnetic emission from the surface of neutron stars in binaries or in isolation~\cite{Miller:2016pom,Ozel:2016oaf,Degenaar:2018lle,Miller:2019cac,Riley:2019yda}; 
as well as information from terrestrial experiments~\cite{Danielewicz:2002pu,Abrahamyan:2012gp,Horowitz:2012tj,Russotto:2016ucm}. These constraints, 
together with their varying levels of statistical and systematic uncertainties, have been extensively 
compared to each other and used to paint an emerging picture about the properties of neutron stars, see e.g.~\cite{Lattimer:2015nhk}. 

The subject of this article is the latest addition to the astrophysical constraints that 
can be used as probes of neutron star properties, 
namely its tidal response. 
When a neutron star is placed in a perturbing tidal gravitational field its shape will
be distorted, expressed through an induced quadrupole moment. Compact binaries that are detectable with gravitational waves provide
a natural stage for this interaction: the neutron star binary component is subject to the gravitational field generated by its companion compact star.
The induced quadrupole moment of the neutron star will, in turn, affect the binding energy of the system and
 increase the rate of emission of gravitational waves~\cite{PhysRev.131.435,PhysRev.136.B1224,Blanchet:2014zz}. The result is a binary system that emits more energy and 
 evolves faster towards the inevitable collision and merger~\cite{Flanagan:2007ix,Hinderer:2007mb}.
 Overall, the tidal properties of the neutron star have a direct imprint on the emitted gravitational wave signal. 
 Detection of the latter can be used to constrain the
 former, a prospect that has received considerable attention over the last few decades~\cite{Faber2012,Baiotti:2016qnr,Baiotti:2019sew}.

This prospect was first realized approximately three years ago, on August 17, 2017, with the detection of gravitational 
waves from a neutron star binary by the second-generation
ground-based detectors LIGO~\cite{TheLIGOScientific:2014jea} and Virgo~\cite{TheVirgo:2014hva},
an event known as GW170817~\cite{TheLIGOScientific:2017qsa}.
The binary source of GW170817 consisted of two neutron stars with masses around $1.35M_{\odot}$ that 
merged approximately $40$Mpc from Earth, resulting in the loudest confirmed gravitational wave signal 
detected to date, as well as the only confirmed detection to unambiguously containing at least one neutron star as of the time of writing~\cite{2018arXiv181112907T}. 
The transient signal spent approximately two minutes in the sensitive frequency band of the detectors, increasing in frequency from $\sim 23$Hz until its merger and accumulating 
$\sim4200$ phase cycles~\cite{Abbott:2018wiz}. 

A second signal likely emitted during a neutron star binary coalescence, known as GW190425, was later detected but it is much weaker than GW170817 as it originated from a much greater distance~\cite{Abbott:2020uma}. 
The masses of the binary components were estimated to be $1.4-1.9M_{\odot}$ ($1.1-2.5M_{\odot}$) when assuming small (arbitrary) spins, 
making this binary not only more massive that GW170817, but also more massive than all known double neutron star 
systems in the Galaxy~\cite{Tauris:2017omb}. The individual component masses are consistent with known neutron star masses
~\cite{Alsing:2017bbc}, however the presence of neutron stars in the binary cannot be established beyond doubt.
If indeed the outcome of a neutron star binary coalescence, though, GW190425 confirms that massive neutron stars form
binaries and merge, somewhat in tension with galactic observations.

After the inevitable merger of the two neutron stars in a binary such as GW170817 or GW190425, a final remnant object is formed.
The gravitational wave signal from the remnant of GW170817 seems to have unfortunately been lost in the detector 
 noise~\cite{Abbott_2017,Abbott:2018hgk,Oliver:2018gbu}. Nonetheless, a potential future detection of such a post-merger signal~\cite{Torres-Rivas:2018svp}
will offer yet another macroscopic probe of neutron star physics, e.g.~\cite{bauswein:15,Baiotti:2016qnr}.
Measurement of the frequency modes of a potential neutron star remnant 
can be used as a source of information that is complementary to the pre-merger constraints on the tidal 
deformability~\cite{Clark:2014wua,Clark:2015zxa,Bose:2017jvk,Chatziioannou:2017ixj,Easter:2018pqy,Tsang:2019esi}
or offer hints of a strong phase transition in the equation of state~\cite{Most:2018eaw,Bauswein:2018bma}.

Besides being the loudest gravitational wave signal at the time of its observation, the source of GW170817 was also
observed with electromagnetic radiation across the spectrum, from gamma rays to radio e.g.~\cite{Monitor:2017mdv,GBM:2017lvd,Troja:2017nqp,Haggard:2017qne,Hallinan:2017woc}, confirming the link between
neutron star coalescences, gamma ray bursts, and heavy element production.
Combined gravitational and electromagnetic data have been extensively used since the detection to not only probe the structure 
of neutron stars -the subject of the rest of this article-, but also to measure the Hubble constant~\cite{Abbott:2017xzu,Hotokezaka:2018dfi} 
and to probe the properties of strong-field gravity~\cite{Monitor:2017mdv,Abbott:2018lct}. To date, no confirmed detection of
counterpart electromagnetic radiation has been reported for GW190425, potentially owing to its higher mass and larger
distance~\cite{Abbott:2020uma,Foley:2020kus,Barbieri:2020ebt,Carracedo:2020xhd}.

Any post-2017 discussion of the tidal deformation of neutron stars and its implications for equation of state constraints will unavoidably be dominated by 
GW170817 and its implications. The detection of GW170817 brought together many different physics communities and served as the catalyst for considerable progress 
in the field in the intervening years. Novel techniques and insights were combined with past experience and results 
to conceive analyses that were applied to the GW170817 and GW190425 data, and that will undoubtedly be again invoked 
on future detections of binary systems involving neutron stars. In this review, we will 
discuss the general properties of the tidal deformability and of tidally-affected gravitational waves from a binary of neutron stars. We will then turn to
GW170817 -and GW190425 to a lesser extent given its weakness- which will serve as a ``shining example" for how a measurement of the tidal deformability from neutron star coalescences can be 
used as a probe of neutron star physics.

For the rest of the article we restrict to geometrized units where $G=c=1$.

%%%%%%%%%%%%%%%%%%%%%%%%%%%%%%%
\section{The tidal deformation of a compact object}
\label{lambda}

Any extended body that is placed in a spatially inhomogeneous external field will experience different forces throughout
its extent. The result is a tidal interaction, the theory of which is well-understood for Newtonian celestial bodies, 
see~\cite{PW} for a modern presentation. In this section we introduce the tidal deformability parameter of a 
neutron star and briefly discuss the corresponding properties of other compact objects such as black holes
and exotic compact objects known as black hole mimickers.

%%%%%%%%%%%%-------------------------------------------------
\subsection{The neutron star tidal deformability}

The tidal deformability of a neutron star is a single parameter $\lambda$ that quantifies how easily the star is deformed 
when subject to an external tidal field. In general, a larger tidal deformability signals a larger, less compact star that is easily
deformable. On the opposite side, a star with a smaller tidal deformability parameter is smaller, more compact, and it is more difficult to deform. 
Mathematically, for both Newtonian and relativistic stars the tidal deformability is defined as the ratio of the induced quadrupole $Q_{ij}$ to the perturbing tidal field ${\cal{E}}_{ij}$ that causes the perturbation 
\begin{equation}
\lambda \equiv -\frac{Q_{ij}}{{\cal{E}}_{ij}}.\label{lambdaref}
\end{equation}

The above equation and dimensionality arguments suggest that $\lambda$ is a sensitive function of the radius of the neutron star. 
In the Newtonian limit the perturbing tidal field ${\cal{E}}_{ij}$ is defined as the second spatial derivative of the external field,
resulting in units of inverse length squared~\cite{PW}. The quadrupole moment has units of length cubed. 
Overall, we expect $\lambda \sim \kappa R^5$, where $\kappa$ is a dimensionless constant and $R$ is the radius of the neutron star,
which sets the length scale of the system. Keeping up with traditional conventions, we express the tidal deformability as
\begin{equation}
\lambda  = \frac{2}{3} k_2 R^5, 
\end{equation}
where $k_2$ is the gravitational Love number with typical values around $0.2-0.3$ for different equations of state. The internal structure of
the neutron star is imprinted on the tidal deformability $\lambda$ 
through both the Love number $k_2$ and the radius of the star $R$. In that sense the tidal deformability contains additional information
compared to pure radius measurements and can be thought of as a complementary probe of neutron star structure.

A related quantity that is commonly also used is the dimensionless tidal deformability defined as
\begin{equation}
\Lambda \equiv \frac{\lambda}{m^5} = \frac{2}{3} k_2\frac{R^5}{m^5} = \frac{2}{3} k_2 C^{-5}\label{Lambdadim},
\end{equation}
where $m$ is the mass of the star, and $C\equiv m/R$ is its compactness. In the following, we will use the terms ``tidal deformability" and 
``dimensionless tidal deformability" interchangeably. 

%%%%%%%%%%%%-------------------------------------------------
\subsection{Computing the tidal deformability}

Ignoring the star's rotation, a single unperturbed neutron star will be static and spherically symmetric. 
By Birkhoff's theorem, the exterior spacetime will be given by the Schwarzschild metric.
If placed under the influence of an external tidal field, however, the star will be deformed and the spacetime will be affected 
accordingly. In the local asymptotic rest frame of the star and in mass-centered Cartesian coordinates
the time-time component of the metric can be expressed as~\cite{PhysRevD.58.124031} 
\begin{align}
\frac{1-g_{tt}}{2}&=-\frac{m}{r}-\frac{3Q_{ij}}{2r^3}\left(n^i n^j-\frac{1}{3}\delta^{ij}\right)+O(r^{-4})\nn\\
&+\frac{1}{2}{\cal{E}}_{ij}x^ix^j+O(r^3),\label{metricpert}
\end{align}
for large distances to the star $r$, and where $x_i$ are the coordinates and $n_i=x_i/r$. 
In the first line of the equation, the first term corresponds to the usual monopolar field of a star with mass $m$, 
the second term corresponds to the quadrupolar term in a field 
expansion which in this case originates by the body's response to the tidal field, 
and $O(r^{-4})$ terms are related to potential higher order multiple moments.
In the second line, the first term is the external tidal contribution
and $O(r^3)$ terms are caused by potential higher order external tidal fields. 
The perturbed metric in Eq.~\ref{metricpert} effectively
defines both the tidal field and the resulting quadrupole moment, see~\cite{GuerraChaves:2019foa} for a detailed discussion.

Given a proposed equation of state, the above metric ansatz, and the definition of Eq.~\eqref{lambdaref}, 
the tidal deformability of a neutron star of a certain mass can be obtained by computing
the metric in the asymptotic regime using the 
Einstein equations, extracting the corresponding $r$ order terms, and taking their ratio.
The ensuing calculation is described in detail in~\cite{Hinderer:2007mb}, in which Hinderer obtains an expression for 
$\lambda$ (or equivalently $k_2$) by imposing continuity of the metric and its derivatives across the 
surface of the neutron star and expanding the metric solution asymptotically~\cite{1967ApJ...149..591T}.
The result is an expression for the gravitational Love number $k_2$ in terms of the value of a metric function and its derivative on the star's surface
\begin{equation}
k_2=k_2(C,y),\label{k2comp}
\end{equation}
where $y$ is a quantity that depends on the value of a metric function and its derivative on $R$, 
and $C$ is the compactness of the star. 
Hinderer also finds that a fully relativistic computation of $k_2$ such as the above results in 
differences around $10-20\%$ compared to the Newtonian results~\cite{Hinderer:2007mb}.

\begin{figure}[]
\includegraphics[width=\columnwidth,clip=true]{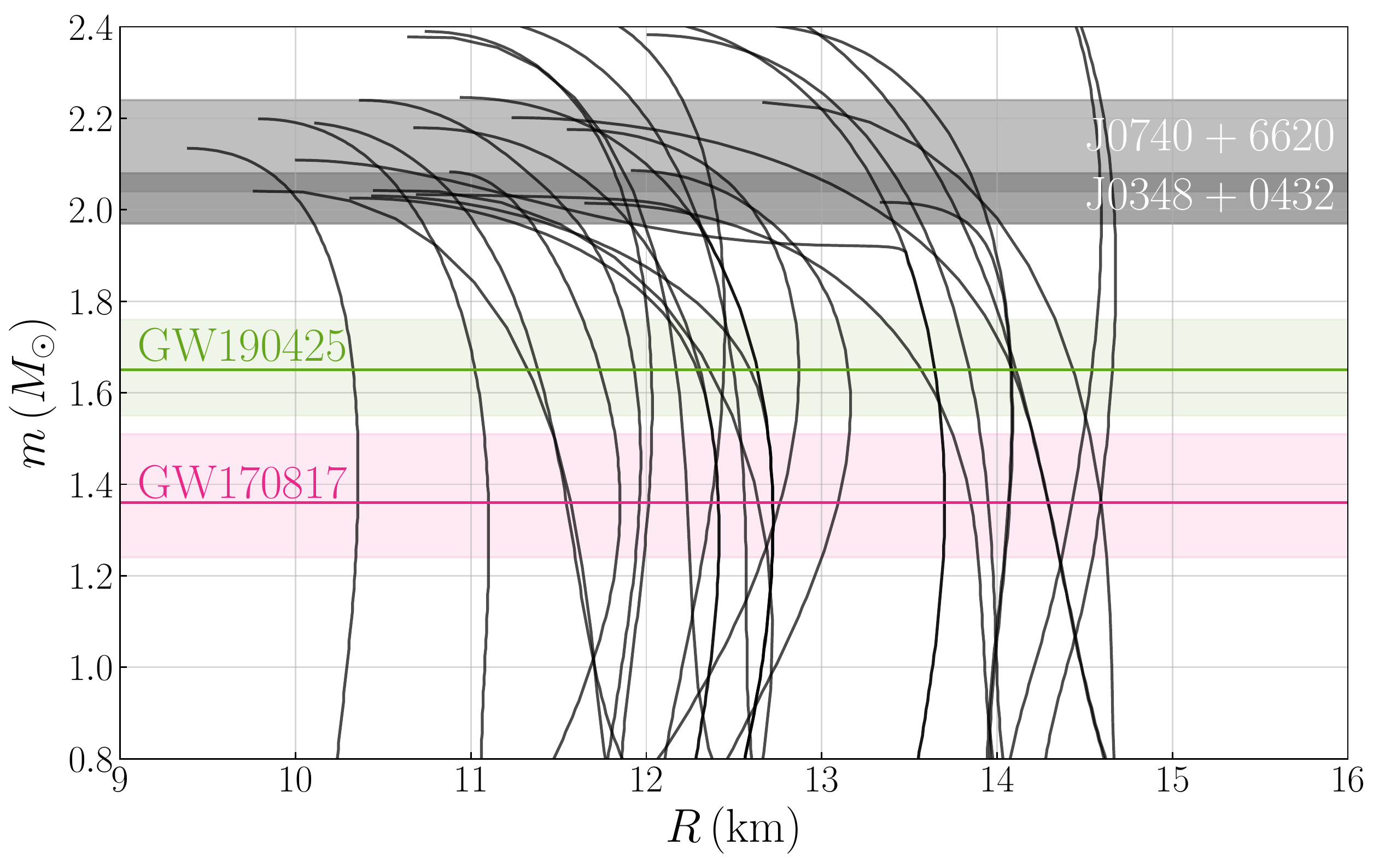}
\includegraphics[width=\columnwidth,clip=true]{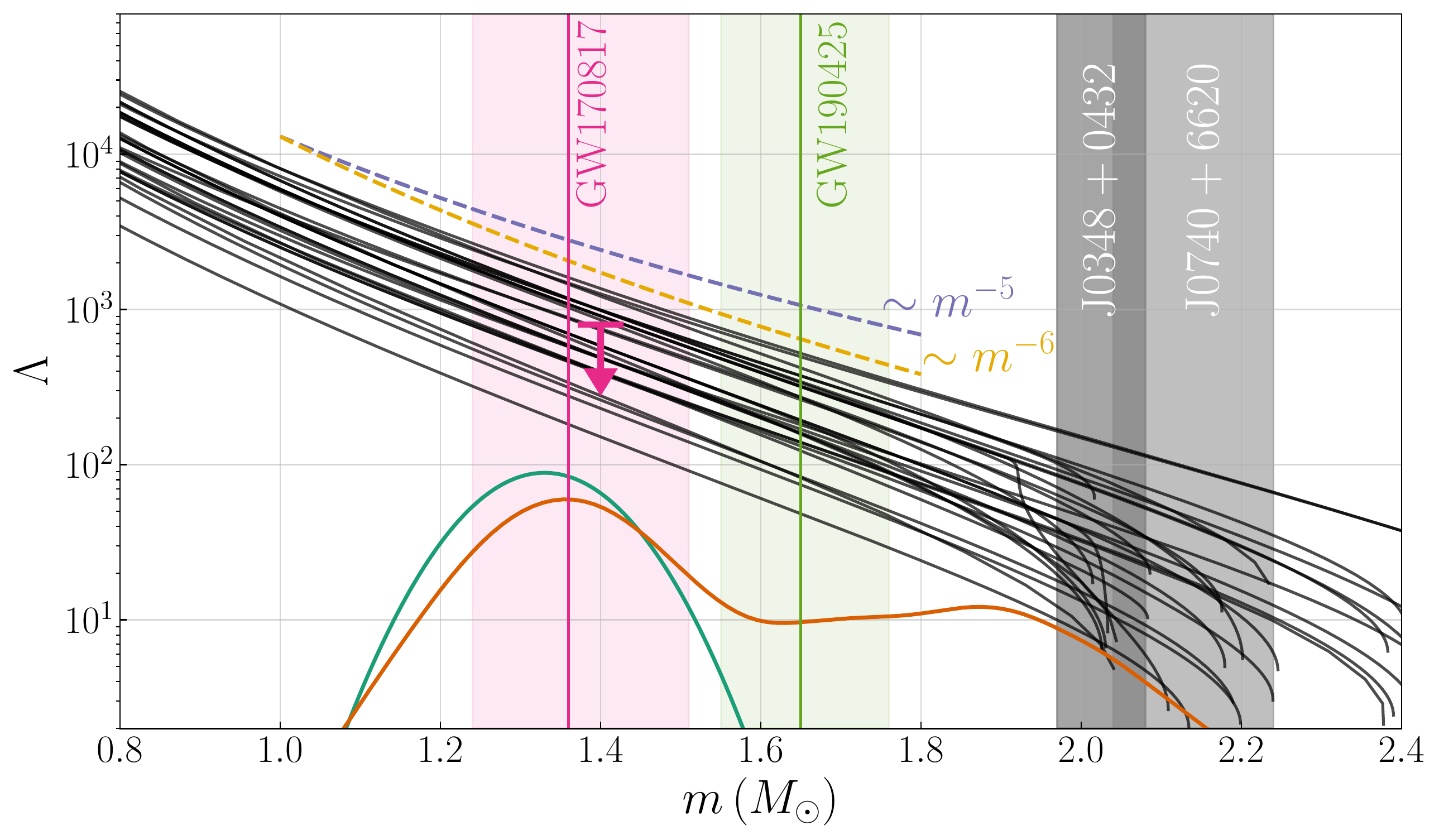}
\caption{Neutron star mass $m$ as a function of radius $R$ (top panel) and dimensionless tidal deformability $\Lambda$ as a
function of mass $m$ (bottom panel) for a wide variety of equation of state models. 
Dashed lines show the $\sim m^{-5}$ and $\sim m^{-6}$ trend.
The grey shaded bands show the 
1-$\sigma$ mass measurement for J0348+0432~\cite{Antoniadis:2013pzd} and J0740+6620~\cite{Cromartie:2019kug}
which serve as lower limits on the maximum neutron star mass possible. The pink and green shaded regions are the 1-$\sigma$
constraint on the binary component masses from GW170817~\cite{Abbott:2018wiz} and GW190425~\cite{Abbott:2020uma} respectively, obtained assuming that the dimensionless spins of the neutron stars are below $0.05$. Solid vertical lines of the same color
denote the ``equal mass" limit, i.e. the masses of the two binary components if assumed equal and given the binary chirp mass. 
For each signal, one binary component mass is larger and one is smaller than the value denoted by the vertical line.
The pink arrow denotes the $\Lambda_{1.4}<800$ upper limit derived from GW170817~\cite{TheLIGOScientific:2017qsa}.
The green curve is a fit to the masses of galactic double neutron star systems~\cite{Ozel:2016oaf}, while the orange line is the 
bimodal fit to the masses of galactic neutron stars~\cite{Antoniadis:2016hxz,Alsing:2017bbc}, expressed through the marginalized
population distribution from~\cite{2020arXiv200500032F}. The height of the green and orange fit distributions is arbitrary.
}
\label{fig:EoSs}
\end{figure}

Figure~\ref{fig:EoSs} shows the neutron star mass as a function of the radius and the dimensional tidal deformability as a function
of the mass for various equation of state models of varying stiffness. For reference, we also show the mass estimates for the
two likely most massive radio pulsars with precise mass measurements~\cite{Antoniadis:2013pzd,Cromartie:2019kug}, 
as well as mass estimates for GW170817~\cite{Abbott:2018wiz} 
and GW190425~\cite{Abbott:2020uma}. 
The two curves on the bottom panel (arbitrary height) correspond to the mass distribution of galactic
neutron stars observed electromagnetically. The green curve is a fit to double neutron star systems~\cite{Ozel:2016oaf},
while the orange bimodal one also includes neutron stars in X-ray and white dwarf-neutron binaries~\cite{Antoniadis:2016hxz,Alsing:2017bbc}.

The bottom panel of Fig.~\ref{fig:EoSs} shows that the dimensionless tidal deformability $\Lambda$ is a steep, monotonically
decreasing function of the mass, covering many orders of magnitude. Indeed, less massive neutron stars are less compact, 
and hence more easily deformable, quantified through a larger $\Lambda$ value than more massive stars.
Dashed lines are proportional to 
$m^{-5}$ and $m^{-6}$, suggesting that $\Lambda$'s dependence on the neutron star mass is closer to being inversely
proportional to the sixth rather than the fifth power. The additional factor of $m$ compared to the expectation from
 Eq.~\ref{Lambdadim} comes from the fact that the tidal Love number $k_2$ is also approximately inversely proportional to the
  mass in the relevant mass range~\cite{Zhao:2018nyf}. For a given equation of state, $\Lambda$ 
  varies from ${\cal{O}}(10^4)$ for $1M_{\odot}$ stars, to ${\cal{O}}(10)$ for the most massive neutron stars possible.
  Additionally, at a fixed neutron star mass different equation of state models predict values for $\Lambda$ spanning about one
  order of magnitude. This dramatic dependence of $\Lambda$ on both the neutron star mass and the equation of state
  is what makes it a promising probe of the neutron star internal structure, as discussed in Sec.~\ref{sec:phase}.

%%%%%%%%%%%%%-------------------------------------------
\subsection{Beyond the tidal deformability}

The tidal deformability $\lambda$ and its associated Love number $k_2$ are the most commonly considered tidal parameters in the context of 
binary neutron star systems. They are, however, only the first in a series of tidal parameters of different types and orders
that arise from the study of generic tidal perturbations. 
The precise, gauge-invariant formalism of tidal Love numbers in full General Relativity and for arbitrarily strong gravitational fields
was developed in parallel in~\cite{Damour:2009vw,Binnington:2009bb}. 

Both studies considered the electric-type and magnetic-type
Love numbers as well as higher multiple moment orders beyond the quadrupole.
The electric-type Love numbers arise from the response of a star to the external electric tidal field. In Newtonian gravity the electric tidal field is defined through partial
spatial derivatives of the external gravitational potential. Higher order derivatives of the external field correspond to higher multiple order electric tidal fields, which are in turn
connected to higher induced multiple moments of the star through higher multiple order Love numbers. For example, $k_2$ is the quadrupole Love number
which relates the induced quadrupole moment to the external quadrupole tidal field $Q_{ij}\sim k_2 R^5{\cal{E}}_{ij}$, obtained after two differentiations of the external field. By analogy, $k_3$ 
is the octupole Love number and it relates the induced octupole moment to the external octupole tidal field $Q_{ijk}\sim k_3 R^7{\cal{E}}_{ijk}$, which is now obtained after three 
differentiations of the external field. Higher order terms and multiple moments can be defined similarly.
The magnetic-type Love numbers arise from the response of a star to the external magnetic tidal field, which corresponds to odd-parity 
sector perturbations. The magnetic-type Love numbers have no Newtonian analogue. 

Besides higher order corrections, the tidal deformability $\lambda$ also ignores the dynamical
response of the tidal field~\cite{Steinhoff:2016rfi} and the effect of the star's rotation. 
The former should be subdominant to the adiabatic effect quantified by $\lambda$ for neutron star coalescences
as the internal neutron star deformation timescale is faster than the orbital motion.
For the latter, additional ``rotational" Love numbers have been
introduced to express the coupling between the external tidal field 
and the star's spin angular momentum~\cite{Landry:2015zfa,Pani:2015hfa,Pani:2015nua,Landry:2017piv,Abdelsalhin:2018reg}. The influence of the rotational Love numbers
on the gravitational wave signal emitted during a neutron star coalescence will depend on the spin of the neutron stars, which might be small if the 
known galactic binary systems are any indication~\cite{Tauris:2017omb}.

Current analyses of gravitational wave signals from neutron star binaries focus on the leading-order electric tidal deformability $\lambda$.
With improving detectors, though, it might be possible to measure higher-order effects. Reference~\cite{Jimenez-Forteza:2018buh} showed that
the magnetic tidal Love numbers could be within reach of next generation detectors~\cite{2010CQGra..27h4007P,2011CQGra..28i4013H,ISwhitePaper}.
The spin-tidal couplings, on the other hand, only become relevant for relatively highly-spinning neutron stars with a dimensionless spin magnitude greater
than $\sim 0.1$~\cite{Jimenez-Forteza:2018buh}
(a dimensionless spin of $\sim 0.4$ corresponds to a rotational period of $\sim 1$ms).
For typical binary neutron star systems observed by second-generation ground-based detectors in the next few years, the influence of higher-order tidal corrections beyond 
$\lambda$ is expected to be negligible.

%%%%%%%%%%%%%%-----------------------------------------
\subsection{The tidal deformability of a black hole}

Besides the precise formalism of tidal deformabilities for relativistic neutron stars, Refs.~\cite{Damour:2009vw,Binnington:2009bb} also address the issue of
the tidal properties of nonrotating black holes. Both studies state that even though the black hole horizon can be deformed away from
spherical symmetry, the tidal deformability of a non rotating black hole $\lambda_{\mathrm{BH}}$ is identically zero. 
Binnington and Poisson~\cite{Binnington:2009bb} argue that this is the 
correct and self-consistent interpretation of their results, as $\lambda_{\mathrm{BH}}=0$ is the only condition that can lead to a unified treatment of neutron stars and black holes. 
Their formalism involves expressing the Schwarzschild metric of an unperturbed nonrotating body in 
geometrically-meaningful light-cone coordinates, in which the advance-time coordinate is constant along past light cones that converge towards $r=0$.
This metric applies to both neutron stars and black holes whose exterior spacetime is identical when static and spherically symmetric, resulting 
in a unified treatment of unperturbed compact objects.

Once the spacetime is perturbed along the lines of Eq.~\eqref{metricpert} and~\cite{PhysRevD.58.124031}, the only choice  
that can maintain the notion of a unified treatment is for the tidal deformability of back holes to be zero. Indeed the resulting 
perturbed spacetime metric diverges for $r=2m$, i.e. on the event horizon of a nonrotating black hole, unless the tidal deformability of the object vanishes identically. This concern is of course moot for neutron stars, 
whose surface is always outside $2m$, making their exterior spacetimes regular everywhere even for non zero tidal deformabilities.
A zero tidal deformability for black holes then results in a perturbed spacetime
that is regular in the relevant exterior region and can describe both black holes and neutron stars. 

Gralla challenged this reasoning by arguing that a unified treatment of perturbed black holes and neutron stars is not a unique choice~\cite{Gralla:2017djj}.
He instead proposed that the relevant physical quantity that is unambiguously defined and accessible to gravitational wave measurements is the \emph{difference} between the tidal deformabilities 
of neutron stars and black holes, i.e. $\lambda_{\mathrm{NS}}-\lambda_{\mathrm{BH}}$. 
As briefly discussed in the next section, and in detail in~\cite{Dietrich:2020eud}, this difference is also in practice what we estimate with gravitational wave measurements.
The waveform models with which we analyze neutron star binaries are tidally-enhanced versions of existing black hole binary models that have been
calibrated against numerical relativity simulations. Therefore both the black hole and neutron star tidal deformability is accounted for and what we measure is 
the difference between the two effects.
 
However, the distinction between the tidal deformability of a neutron star and its difference to that of a black hole might be more important when comparing measurements to theoretical expectations.
For a given equation of state model, Eq.~\eqref{k2comp} can be used to compute the corresponding expected Love number which can in turn be compared to the measured
tidal deformability from gravitational wave data. Any ambiguity in the interpretation and comparison between the measured quantity and the one estimated through perturbation theory with Eq.~\eqref{k2comp}
  could result in a systematic error.
Gralla estimates that the size of the ambiguity is less than $10$ in the dimensionless tidal deformability~\cite{Gralla:2017djj}, which is beyond the reach of current 
detectors but could become important in the future, if indeed present.

Concluding this discussion, it is interesting to note that the tidal deformability can be used as a probe of non standard physics.
For example, ``black hole mimickers" 
-exotic compact objects that are similar to black holes- might have non negligible, or even negative, tidal deformabilities~\cite{Cardoso:2017cfl}. 
If the tidal deformability of black holes is indeed vanishing (or if it is very small), studying tidal effects in mergers of black holes could 
 put constraints on horizon-level spacetime corrections~\cite{Addazi:2018uhd}. The tidal deformability can therefore also be used to
test the nature of the coalescing compact objects~\cite{Wade:2013hoa,Johnson-McDaniel:2018uvs,Maselli:2017cmm,Chirenti:2020bas,Fransen:2020prl},
or other physics such as dark matter and extra dimensions~\cite{Nelson:2018xtr,Quddus:2019ghy,Chakravarti:2019aup,Zhang:2020pfh}.

%%%%%%%%%%%%%%%%%%%%%%%%%%%%%%%
\section{Measuring the tidal deformability with neutron star coalescences}
\label{BNS}

Neutron star binaries are a natural stage for tidal interactions between the binary components.
In the early stages of the coalescence, the two stars are slowly inspiraling towards each other, emitting
energy in gravitational waves. In this early inspiral stage the two stars are sufficiently
far apart that tidal interactions are negligible and the emitted signal is indistinguishable from that of two
inpiraling black holes of the same masses and spins. Eventually, energy loss will drive the neutron stars
close enough that tidal interactions will become important. As described in the previous section, each neutron star
will be subject to the gravitational tidal field produced by its companion. As a result, its shape will be adiabatically 
deformed: the time scale associated with this deformation is faster than the orbital motion that modulates the external
tidal field, so the tidal bulge in each star faces the companion.

The induced tidal quadrupole moment on the two binary components will affect the subsequent evolution of the system.
Gravitational wave emission is sourced, to leading order, by a time-varying 
quadrupole moment~\cite{PhysRev.131.435,PhysRev.136.B1224,Blanchet:2014zz}. For a binary, the leading-order contribution
in the quadrupole moment corresponds to the binary orbiting motion. For neutron star binaries, however,
the induced tidal quadrupole moment on the shape of each star results in a subleading contribution that has a direct imprint on the
gravitational wave signal. 
In this section, we quantify this contribution and describe how it can be used to constrain
the tidal properties of neutron stars.

%%%%%%%%%%%%-------------------------------------------------
\subsection{The gravitational wave phase}
\label{sec:phase}

Tidal effects are expected to impact both the amplitude and the phase of the emitted signal, however the latter
is typically better measured for harmonic functions such as waves.
Indeed, the GW170817 signal accumulated $\sim $ 4200 gravitational wave cycles between 23Hz and
merger~\cite{Abbott:2018wiz}, while its amplitude increased by a factor of only about 10.
During the inspiral phase of the coalescence, the (time derivative of the) binary orbital 
frequency can be obtained from the binding
energy of the system and the rate of energy emission through 
\begin{equation}
\frac{d F}{d t} = \frac{d E}{dt}\frac{dF}{d E},
\end{equation}
where $d E/dt$ is the rate of energy emission due to gravitational waves 
and $dF/d E$ is the derivative of the frequency with respect to the binding energy; it can be obtained through Kepler's law.

Both the energy loss and the binding energy are typically expressed as a post-Newtonian 
expansion~\cite{Blanchet:2014zz}, a series expansion in small binary velocity $u$ compared to the speed of light.
A term proportional to $u^{2N}$ relative to the leading order term is referred to as an $N$PN contribution.
In this framework, the leading-order binary quadrupole moment (i.e. a Newtonian quadrupole created by the binary motion) scales as $\sim r^2$
where $r$ is the separation between the two bodies.
The tidal correction to the quadrupole scales as $\sim 1/r^3$, suggesting that 
the tidal part is a relative $\sim1/r^5\sim u^{10}$ -also known as 5PN-
correction compared to the leading Newtonian term.
The above shows why tidal effects are negligible when the stars' separation is large, but become more important as the binary approaches the later stages of its coalescence where velocities increase.

The ensuing calculation results in the following expression of the gravitational wave phase
in the frequency domain for quasicircular inspirals~\cite{Blanchet:2014zz,Flanagan:2007ix}
\begin{align}
\Psi(f)&=2 \pi f t_c + \phi_c -\frac{\pi}{4} \nn
\\
&+ \frac{3}{128 \eta u^5}\left\{1+ f(\eta)u^2 +
(4\beta-16\pi)u^3+ \left[g(\eta)+\sigma \right] u^4+...\right.\nn
\\
&\left.-\frac{39}{2}\tilde{\Lambda}u^{10}+...\right\}.\label{phase}
\end{align}
In the above equation $f$ is the gravitational wave frequency, $t_c$ is the time of coalescence, $\phi_c$ is the phase of coalescence, $\eta\equiv m_1 m_2/M^2$ is the reduced mass, $u\equiv (\pi M f)^{1/3}$, 
$M=m_1+m_2$ is the total mass of the binary, and $m_1>m_2$ are the component masses. The mass dependence of
the leading order term can be expressed as $\eta M^{5/3} = (M \eta^{3/5})^{5/3}$, leading to the definition of the chirp mass
${\cal{M}}\equiv M \eta^{3/5}$, the best measured mass parameter.
The functions $f(\eta)$ and $g(\eta)$
can be found, for example, in~\cite{Blanchet:2014zz}, 
while the terms $\beta$ and $\sigma$ represent spin-orbit and spin-spin contributions to the phase respectively
and are also available in~\cite{Blanchet:2014zz}.

In the third line the term proportional to $u^{10}$ is the leading-order tidal correction, 
originally computed in~\cite{Flanagan:2007ix} and commonly expressed through the parameter~\cite{Favata:2013rwa}
\begin{equation}
\tilde{\Lambda}\equiv\frac{16}{3}\frac{(m_1+12m_2)m_1^4\Lambda_1+(m_2+12m_1)m_2^4\Lambda_2}{(m_1+m_2)^5},\label{lambdaTdef}
\end{equation}
where $\Lambda_1\equiv \lambda_1/m_1^5$ and $\Lambda_2\equiv \lambda_2/m_2^5$ are the dimensionless
tidal deformabilities of each binary component. The main matter effect in the gravitational wave signal is, therefore,
expressed through the tidal deformability of each binary component.
Indeed, numerical simulations of merging neutron stars suggest that changes in the equation of state that 
are not reflected in direct changes in the tidal deformability are difficult to measure 
through inspiral gravitational wave signals~\cite{Read:2013zra}. If one of the binary components is a black hole, then the corresponding dimensionless tidal deformability should be set to zero.

Despite the relative high post-Newtonian order of the tidal correction, the term is measurable due to the fact that the dimensionless
tidal deformability parameters $\Lambda_1,\Lambda_2$ are large for realistic equation of state models,
as shown in the bottom panel of Fig.~\ref{fig:EoSs}. Schematically they are given by 
$\Lambda \sim k_2 C^{-5}$; the tidal Love number is ${\cal{O}}(0.1)$, 
but the fifth power of the compactness results in typical
values of $\Lambda\sim{\cal{O}}(100)$ for realistic equations of state and neutron stars around $1.4M_{\odot}$~\cite{Hinderer:2009ca}. 
For $\tilde{\Lambda}=400$ (a value consistent with GW170817~\cite{Abbott:2018wiz} and 
nuclear equation of state models) the prefactor of the tidal term is $7800$ times larger than the leading order Newtonian term.
The combination of the large prefactor and the high order of the tidal term suggests that tidal interactions are a small but 
potentially not
negligible effect~\cite{Hinderer:2009ca}.
Interestingly, for some particularly stiff equations of state that result in neutron stars with radii $\sim15$km, 
ignoring the tidal term in the gravitational wave phase when searching for these signals could even decrease detection efficiency~\cite{Cullen:2017oaz}.

\begin{figure}[]
\includegraphics[width=\columnwidth,clip=true]{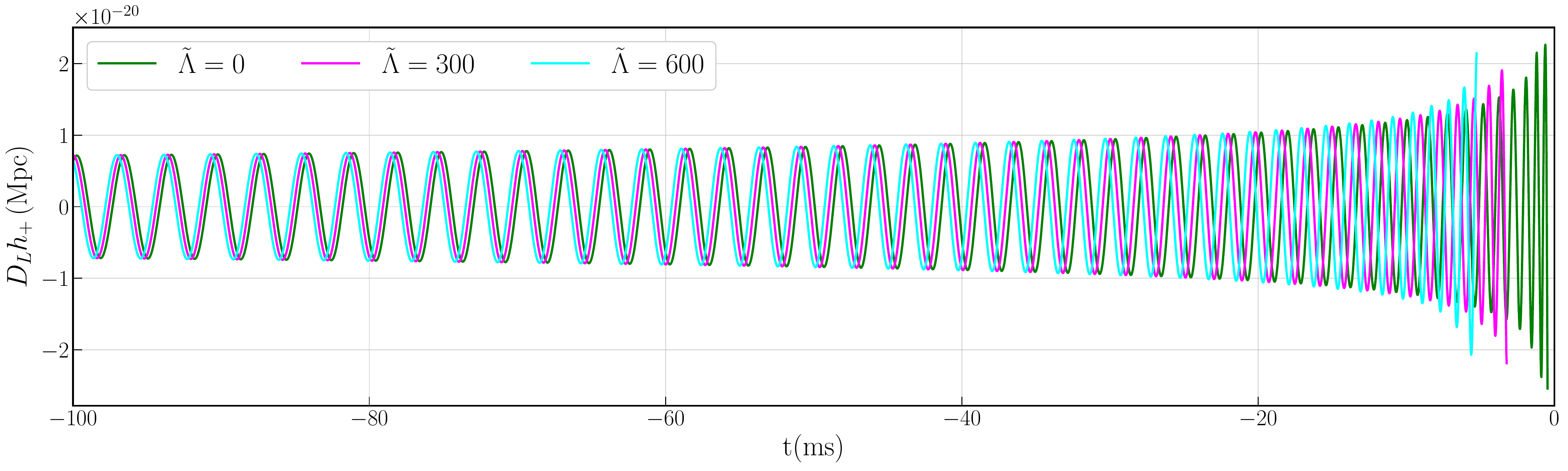}
\includegraphics[width=\columnwidth,clip=true]{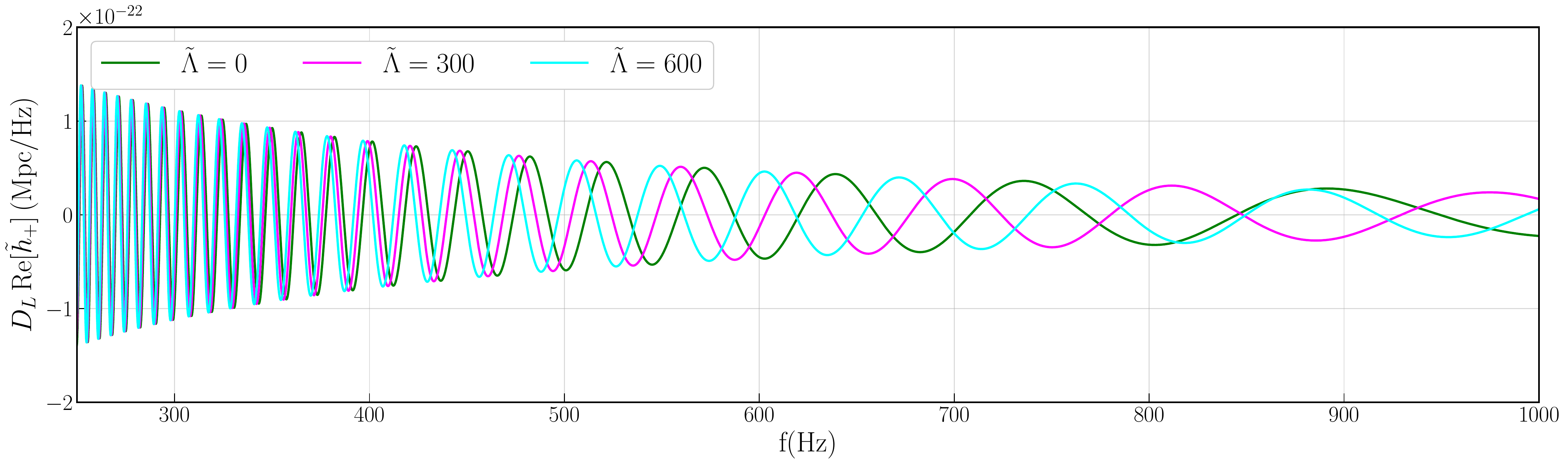}
\caption{Scaled gravitational wave signal from the late inspiral of two equal-mass, non spinning neutron stars for different 
values of  $\tilde{\Lambda}$ using the {\tt IMRPhenomD\_NRTidalv2}~\cite{Dietrich:2018uni,Dietrich:2019kaq} waveform model.
We plot the plus polarization in the time domain in the top panel, and the real part of the plus polarization in the frequency domain in the bottom panel.
Since we employ an analytic model, waveforms have been terminated at the peak of the time domain waveform amplitude,
sometimes used to approximately signify the merger. 
}
\label{fig:waveform}
\end{figure}

The impact of tidal interactions on the waveform evolution is shown in Fig.~\ref{fig:waveform} where we plot the plus polarization
 of the gravitational wave in the time (top) and frequency
 (bottom) domain. We use the {\tt IMRPhenomD\_NRTidalv2}~\cite{Dietrich:2018uni,Dietrich:2019kaq} waveform model and 
 focus on the late stages of the coalescence of two equal-mass, nonspinning neutron stars with $m=1.4M_{\odot}$ for different
 values of $\tilde{\Lambda}$. At early times/low frequencies tidal interactions are weak and the three waveforms are similar. 
 As the two neutron stars approach each other, tidal interactions become stronger, with larger values of $\tilde{\Lambda}$
 leading to enhanced energy emission and a faster overall system evolution. This dephasing between the three waveforms is the
 dominant matter imprint on inspiral gravitational signals. 

The expression for the phase in Eq.~\eqref{phase} does not include a number of subdominant terms, some of which
have been calculated while others are still unknown. Corrections to the tidal term -which would first enter at 6PN, for a relative $u^{12}$ contribution- have been computed in~\cite{Vines:2011ud,Damour:2012yf}, though their contribution is expected to be small,
 making them very difficult to measure with current detectors~\cite{Wade:2014vqa}. Additional terms not included in the expression of Eq.~\eqref{phase} include the already-mentioned higher-order multiple
corrections, magnetic Love numbers~\cite{Damour:2009vw,Binnington:2009bb}, 
rotational Love numbers~\cite{Landry:2015zfa,Pani:2015hfa,Pani:2015nua,Landry:2017piv,Abdelsalhin:2018reg}, 
terms related to the dynamical response to the tidal field~\cite{Steinhoff:2016rfi},
and terms beyond the linear field approximation. The above also ignores any magnetic field contribution in the binary dynamics
and tidal deformability~\cite{Zhu:2020imp},
a reasonable assumption unless the magnetic field is extremely large, 
$B\sim 10^{16}-10^{17}G$~\cite{Giacomazzo:2009mp,Giacomazzo:2010bx,Palenzuela:2015dqa}.

Besides tidal interactions in the binary, the equation of state also affects the
quadrupole-monopole, or self-spin, term which is a 2PN phase correction. This effect is caused by the fact that the 
shape of the neutron star is deformed under its own spin, resulting in a spin-induced quadrupole
moment. The degree of deformation depends on the equation of state of the star; the leading order effect and its first correction have been computed in~\cite{Poisson:1997ha,Bohe:2015ana}.
Despite being formally a lower order term, the self-spin contribution to the gravitational wave phase is smaller 
than the tidal deformability contribution, and can be neglected unless the neutron star 
is rotating significantly~\cite{Harry:2018hke}. 
The spin-induced quadrupole (and the resulting self-spin term) can be approximately calculated given the tidal
deformability of the star in a way that is approximately agnostic about the underlying equation of state, through the
Love-Q relation~\cite{Yagi:2013bca,Yagi:2013awa}. Waveform models can, therefore, include the self-spin term
directly without the need of additional binary parameters~\cite{Agathos:2015uaa,Chatziioannou:2015uea}.

Finally, it is worth remembering that Eq.~\eqref{phase} does not include a number of point-particle terms that are represented
by the ellipsis. Those terms are of lower or the same post-Newtonian order as 
the tidal terms of interest here and they have fully been computed up to the term 
proportional to $u^{7}$~\cite{Blanchet:2014zz}. Though smaller in magnitude than the tidal terms, their absence 
could lead to biases when measuring the tidal effects
from a gravitational wave signal~\cite{Favata:2013rwa,Yagi:2013baa}. 
However, as already mentioned most waveform models in use are constructed from a baseline binary black hole model
on top of which tidal effects have been added. The binary black hole baseline models are typically compared against 
numerical relativity simulations which are used to calibrate free parameters. This calibration results in some of the missing
non-tidal terms being implicitly accounted for, even if their exact form is not fully known.

%%%%%%%%%%%%-------------------------------------------------
\subsection{Tidal constraints and inference on the equation of state}

Understanding of the imprint of tidal interactions on gravitational wave signals has facilitated 
the development of analyses for measuring tidal parameters and interpreting them in the
context of neutron star equations of state, as well as
studying the prospects of constraining tidal effects with gravitational waves. The general picture suggests
that gravitational waves can be used to measure the radii of neutron stars with kilometer-order precision. This precision is
comparable to constraints from electromagnetic observations, but the measurement is subject to entirely independent 
systematic uncertainties. The detection of GW170817 allowed for the direct application of these methods towards 
a measurement of $\tilde{\Lambda}\lesssim 700$ and $R\lesssim 13$km
at the 90\% level. Before delving into the lessons learned from GW170817 in the next section, we begin by describing 
the various challenges faced and how they might be overcome\footnote{The goal of this section is not to provide a historically
or chronologically faithful account of progress in the field. 
The goal, rather, is to provide the logical steps -at least according to the author- 
that can lead from Eq.~\ref{lambdaTdef} for the $\tilde{\Lambda}$ definition to Fig.~\ref{fig:radius} for the equation of state
constraints after GW170817 and GW190425.}.

%--------------------------------------------------------------------------------------------------
\subsubsection{Measurement accuracy for tidal parameters}

The computation of the tidal phase correction given in Eq.~\eqref{phase}~\cite{Flanagan:2007ix} gave the opportunity to study how well the tidal parameters can be measured with gravitational waves. Initial estimates based on the Fisher
matrix approximation~\cite{Vallisneri:2007ev} and tidally-enhanced post-Newtonian
signals of the form of Eq.~\eqref{phase} were promising~\cite{Flanagan:2007ix,Hinderer:2009ca} as also later confirmed with 
effective-one-body waveforms~\cite{Damour:2012yf}. These studies suggested that, even though tidal effects appear at a very
high post-Newtonian order, they can be measured well-enough to differentiate between the predictions of different equations
of state, as the latter can vary by orders of magnitude, see Fig.~\ref{fig:EoSs}.
The Fisher matrix approximates the likelihood function for the gravitational wave data
 with an expansion up to quadratic order around the parameters of the observed system, which is valid in the high signal-to-noise ratio regime, but can result to inaccurate results for weak signals
or subtle effects such as tidal interactions~\cite{Vallisneri:2007ev}.
Despite that, studies that considered the full likelihood through Monte-Carlo-based sampling methods confirmed that 
$\tilde{\Lambda}$ can be measured to within $\sim 600$ at the 2-$\sigma$ level for signals with a signal-to-noise ratio of 30, 
which is comparable to what was later achieved with GW170817~\cite{Wade:2014vqa}.

Going beyond the leading-order analytic phase computed in the post-Newtonian framework, 
numerical simulations of merging neutron stars~\cite{Baiotti:2016qnr} are able to examine signals
closer to merger, however this comes at the expense of restricting to the limited number of simulations available.
Numerical simulations of the late stages of a neutron star coalescence 
produced with different equations of state have been used to show that the signal is distinguishable from point-particle signals and
 future detections could help determining the neutron star radius to an accuracy of $\sim 1$km~\cite{Read:2009yp}. 
These results have been confirmed with more simulations from different numerical
relativity codes and equations of state~\cite{Read:2013zra} and when 
using longer hybrid waveforms which are obtained by stitching together numerical relativity simulations 
and effective-one-body waveforms for coalescing binary neutron stars~\cite{Hotokezaka:2016bzh}.

%--------------------------------------------------------------------------------------------------
\subsubsection{Measuring the tidal deformability and radius of a neutron star}
\label{LNS}

The functional form of the leading order tidal effect, $\tilde{\Lambda}$ in Eq.~\ref{lambdaTdef}, implies that gravitational wave
signals primarily offer information about a combination of the stars' masses and tidal deformabilities. The neutron star masses can be independently constrained from the lower orders in the post-Newtonian expansion of the waveform phase, 
with the binary chirp mass being 
measured to much higher accuracy than the mass ratio.
However, higher order tidal terms that could
in principle allow for an independent measurement of $\Lambda_1$ and $\Lambda_2$ are not expected to be constrained with
current sensitivity detectors~\cite{Wade:2014vqa}. 

The issue of using $\tilde{\Lambda}$ and the masses alone 
to estimate the stars' tidal deformabilities and, in turn, the equation of state can be tackled with two broad strategies. The first
involves imposing constraints between the two tidal deformabilities and the neutron star radii that are approximately insensitive to
the underlying equation of state. The second relies on making use of some 
phenomenological representation of the equation of state itself and will be more extensively discussed in Secs.~\ref{combining} and~\ref{EoSpar}. Both
approaches rely on the inherent assumption that the coalescing bodies are indeed neutron stars, and their equation of state
resembles existing nuclear models. Instead the assumption of the neutron star-black hole coalescence trivially solves the problem
as then $\Lambda_{BH}=0$. This topic will be discussed in Sec.~\ref{sec:nsbh}.

The dependence of the dimensionless tidal deformability of a neutron star on its mass, Fig.~\ref{fig:EoSs} bottom panel, 
is qualitatively similar 
among a wide range of hadronic equation of state models for $m\lesssim 1.8M_{\odot}$. Nuclear models that experience phase
transitions towards strange degrees of freedom in the relevant mass range require separate treatment. But for purely hadronic 
models, the general picture suggests that $\Lambda$ is a monotonically decreasing function of the mass with a slope around
$\sim m^{-5}$, as implied from its functional form in Eq.~\ref{Lambdadim}, or $\sim m^{-6}$ as obtained after taking into account
 dependence of the Love number on the mass as well~\cite{Zhao:2018nyf}. In other words, the 
 tidal parameters of two neutron stars of certain masses  are not completely independent of each other, and $\tilde{\Lambda}$ must reflect that. 
 
\begin{figure}[]
\includegraphics[width=\columnwidth,clip=true]{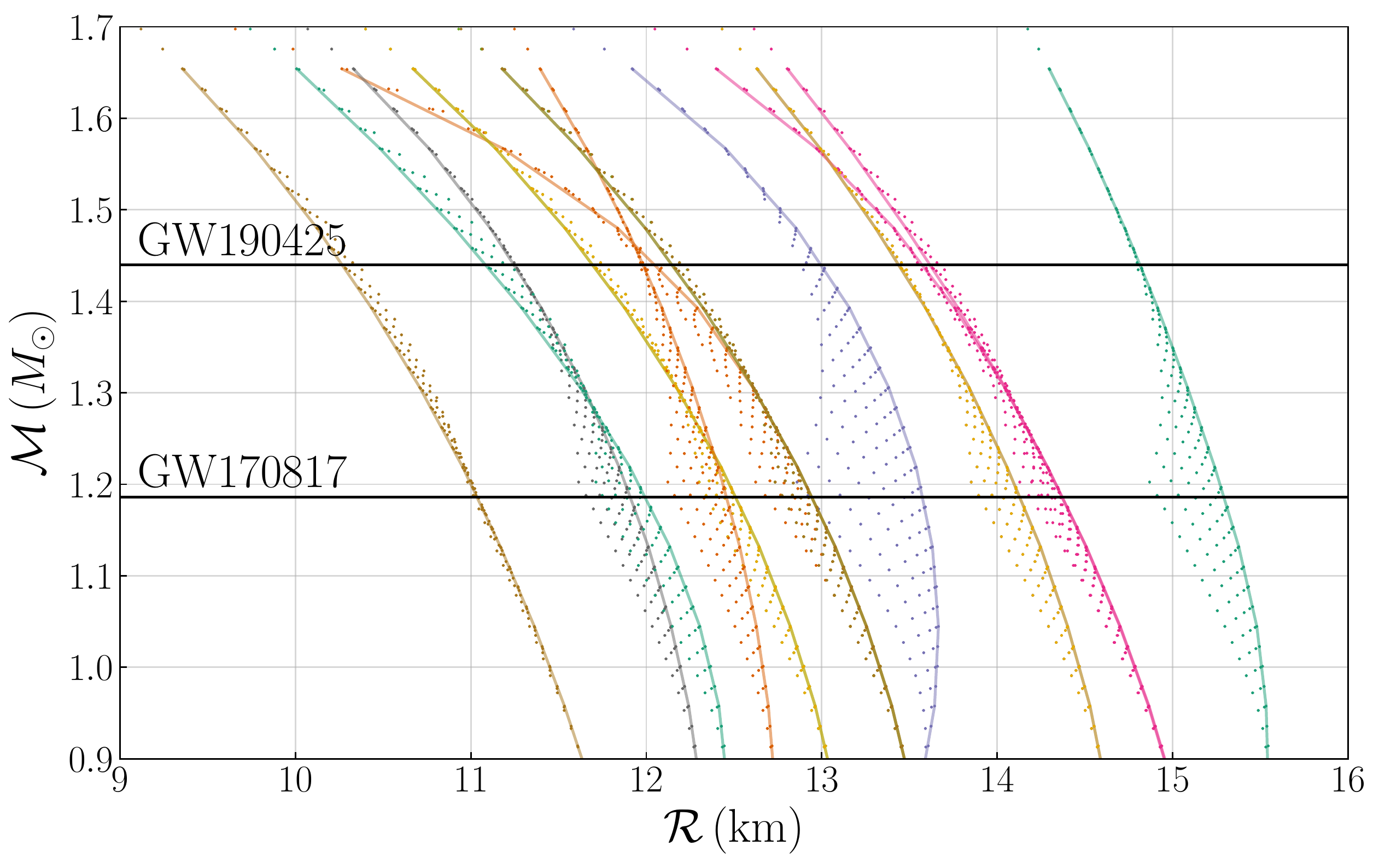}
\includegraphics[width=\columnwidth,clip=true]{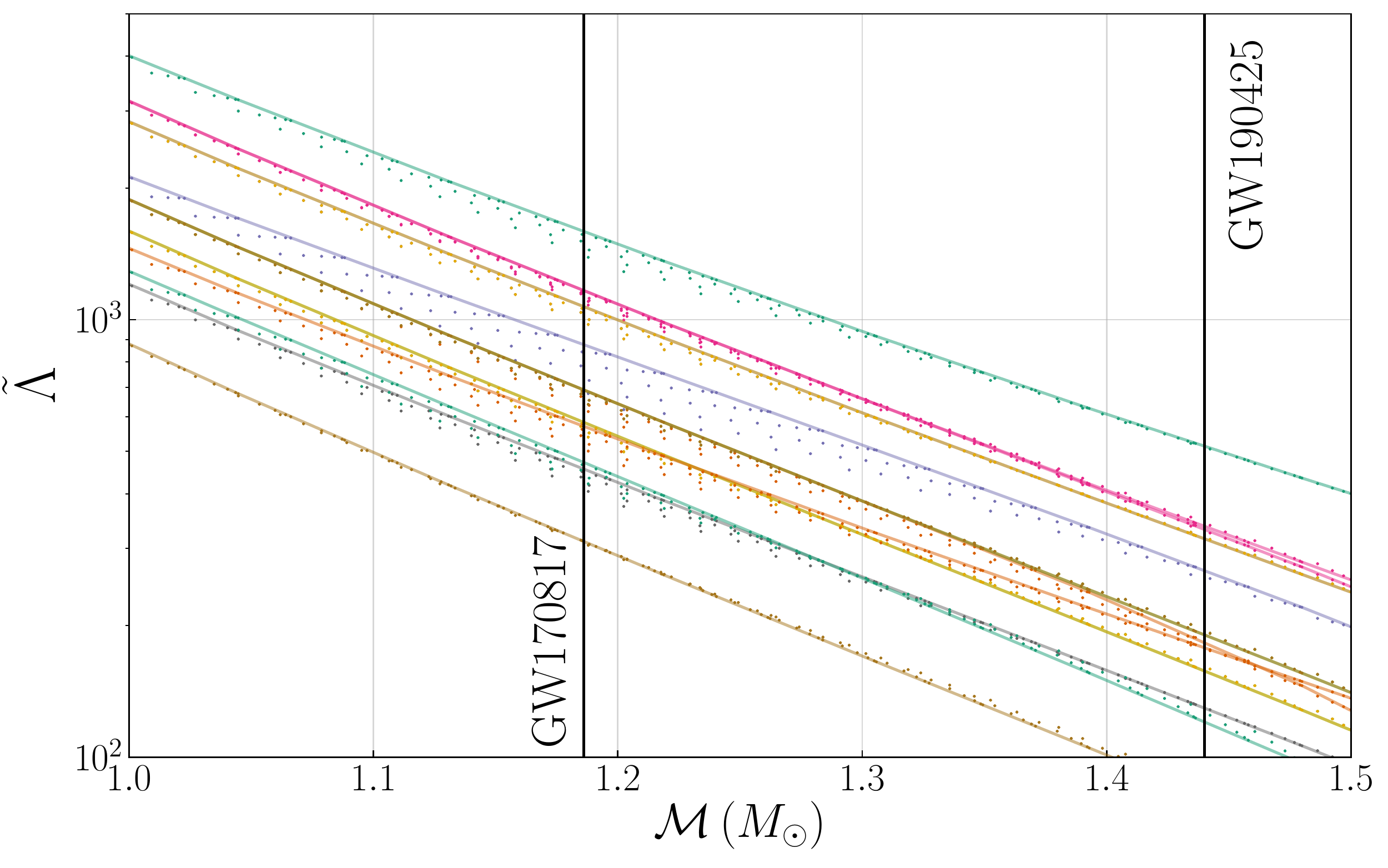}
\caption{Chirp mass as a function of the chirp radius (top panel) and $\tilde{\Lambda}$ as a function of the chirp mass for selected
hadronic equations of state (different colors) and binary mass ratios (dots). For each equation of state we select binaries with 
masses evenly spaced between $1M_{\odot}$ and $2M_{\odot}$ and plot them as dots. Same-color lines correspond to binaries
with equal masses. The horizontal (top) and vertical (bottom) black lines denote the median chirp mass for GW170817 and 
GW190425.
}
\label{fig:universality}
\end{figure}

 Figure~\ref{fig:universality} shows the chirp mass ${\cal{M}}$ as a function of the chirp radius 
 ${\cal{R}}\equiv 2 {\cal{M}} \tilde{\Lambda}^{1/5}$~\cite{Wade:2014vqa} (top) and $\tilde{\Lambda}$
 as a function of the chirp mass ${\cal{M}}$ (bottom) for different hadronic equations of state (different colors). For each equation of state
 binaries with different mass ratios are represented with dots, while lines correspond to the equal mass limit. Both 
 $\tilde{\Lambda}$ and the related quantity ${\cal{R}}$ primarily depend on the typically well-measured binary chirp mass, while
 the mass ratio dependence is subdominant. This simplification was proposed in~\cite{Wade:2014vqa} as a way to interpret 
 a future $\tilde{\Lambda}$ measurement and distinguish between equation of state models. It was later applied on GW170817 
 by~\cite{Raithel:2018ncd} which also provided an analytic justification for this behavior in the Newtonian limit, as well as
 a calculation of the leading-order correction in $\tilde{\Lambda}$ due to the mass ratio. 

Ways to express the above generic qualitative behavior of hadronic equations of state include 
(i) a Taylor expansion of the tidal deformability around a 
fiducial mass~\cite{DelPozzo:2013ala}, (ii) an approximately equation of state insensitive~\cite{Yagi:2016bkt} relation between
the symmetric and antisymmetric tidal deformability combinations for a given binary mass ratio~\cite{Yagi:2015pkc,Chatziioannou:2018vzf,Carson:2019rjx}, and 
(iii) an approximately equation of state insensitive relation of the form $\Lambda_1 = q^{6}\Lambda_2 $ ~\cite{De:2018uhw,Zhao:2018nyf}.
Different relations and approaches have been shown to result in broadly consistent results at least to within
the current statistical errors for tidal 
parameters~\cite{Wei:2018dyy,Bhat:2018erd,Zhao:2018nyf,Chatziioannou:2019yko,Abbott:2020uma}, 
see also Sec.~\ref{170817} for a more detailed
discussion on GW170817 and GW190425. Employing one of the above relations naturally reduces the statistical measurement 
uncertainty for $\Lambda_1-\Lambda_2$ by a factor of $2-10$~\cite{Chatziioannou:2018vzf}, 
though at the expense of a prior assumption that the coalescing 
bodies are neutron stars and the equation of state is hadronic~\cite{Zhao:2018nyf,Carson:2019rjx}. 

Finally, translating bounds on the tidal parameters to bounds on the neutron star radius is also feasible with
similar equation of state insensitive relations. 
The top panel of Fig.~\ref{fig:EoSs} shows the mass-radius relation for different equations
of state, suggesting that in the $1-1.8M_{\odot}$ range, neutron star radii are approximately constant to within $\sim 500$m.
This statement again is valid only for equations of state without a phase transition in the relevant range, and is related
to the requirement that the equation of state must be stiff enough at high densities in order to be able to support
the observed $\sim 2M_{\odot}$ pulsars~\cite{Zhao:2018nyf}. Relevant equation of state insensitive relations include the direct
assumption of a constant radius~\cite{De:2018uhw,Zhao:2018nyf}, a relation between the neutron star tidal deformability and its compactness
$C(\Lambda)$~\cite{Maselli:2013mva}, a ``quasi-Newtonian" relation between $R$ and $\tilde{\Lambda}$~\cite{Raithel:2018ncd},
and a relation between $R_{1.4}$ and $\Lambda_{1.4}$~\cite{Annala:2017llu};
again relations yield broadly consistent results when applied to existing measurements~\cite{Abbott:2020uma}.

%--------------------------------------------------------------------------------------------------
\subsubsection{Combining information from multiple signals}
\label{combining}

The population of neutron star coalescences that will be observed with gravitational waves 
in the coming years will be characterized by an astrophysical mass distribution which is currently unknown. The tidal
effects experienced by each binary will strongly depend on the component masses, which will vary from binary to binary,
and the equation of state, which is expected to be the same throughout the Universe. Combining tidal information 
from multiple signals in order to obtain stronger constraints on the equation of state, therefore requires either identifying
approximately common parameters or hierarchically working with the common equation of state directly.

The approximate equation of state insensitive phenomenology of hadronic equations of state enabled 
Ref.~\cite{DelPozzo:2013ala}
to follow the first approach by approximating the (dimensionfull) tidal deformability through a Taylor expansion around a fiducial 
mass as 
$\lambda(m)=\lambda_{1.4}+\lambda'_{1.4}(m-1.4)$. Each detected binary, then, provides a constraint on $\lambda_{1.4}$,
the tidal deformability of a $m=1.4M_{\odot}$ neutron star, while the slope parameter $\lambda'_{1.4}$ is not constrained by the data. 
By simulating an astrophysical population of binary neutron star coalescences observed by LIGO and
Virgo at design sensitivity and analyzing them with Monte Carlo techniques to draw samples from the posterior distribution
of the binary parameters~\cite{Veitch:2014wba}, Ref.~\cite{DelPozzo:2013ala} showed that a few tens of detections can 
lead to a $\sim 10\%$ 
measurement of $\lambda_{1.4}$ and can be used to distinguish between soft, moderate, and stiff equations of state,
as also confirmed in~\cite{Wang:2020xwn}.
This result is unaffected by the inclusion of further relevant physical effects such as neutron star spins,
the self-spin interaction term, higher-order tidal effects, and different waveform termination conditions~\cite{Agathos:2015uaa}. 

The use of phenomenological macroscopic relations to combine information from signals raises the issue of
systematic errors as the number of detections increases and might be restricted to hadronic equations of state. 
One way to mitigate the above is to instead combine information at the level of the microscopic equation of state
(i.e. the pressure as a function of the density, as coalescing neutron stars are expected to be cold and temperature effects will be
negligible), which is expected to be common among all neutron stars.
This approach is known as ``hierarchical" since the problem is divided in different levels. At a low level we have the
event-specific constraints on relevant parameters, such as the masses and tidal parameters of GW170817. At a higher level, 
each event-specific constraint informs the common equation of state where different signals (or even different messengers,
such as electromagnetic radii constraints) can be combined. 
See Sec.~\ref{hierarchical} for a short introduction to hierarchical inference.
Along the lines of this approach and guided by the desire to both translate macroscopic constraints to microscopic inference, and
to combine different types of macroscopic constraints, a number of generic representations of the equation of state have been
proposed, see Sec.~\ref{EoSpar} for a discussion on such models and parameterizations. The result of the analysis and their 
applicability will inevitably depend on the equation of state representation employed.

A common equation of state representation is the one in terms of piecewise polytropes.
Using a 4-parameter piecewise model~\cite{Read:2008iy} and a hierarchical framework, Lackey and Wade~\cite{Lackey:2014fwa}
 demonstrated that the detection
of multiple neutron star coalescences can not only be used to measure the tidal deformability of neutron stars, but 
also the microscopic equation of state itself. One year of observations with realistic binary neutron star detection rates 
would suffice to measure
the neutron star radius to $\pm 1$km, also confirmed in~\cite{Abdelsalhin:2017cih,Vivanco:2019qnt}. 
The loudest signals observed (signal to noise ratio above $\sim 20$)
seem to provide almost all relevant constraints, something familiar to multiple-parameter inference~\cite{Haster:2020sdh}. 
The arbitrary polytrope slopes of the piecewise model, suggest that it is well suited for equations of state with strong
phase transitions, however it suffers from an increase in statistical error near the joining densities of the piecewise 
polytropes~\cite{Lackey:2014fwa}. Instead, a spectral parametrization of the equation of state
~\cite{Lindblom:2010bb,2012PhRvD..86h4003L,Lindblom:2013kra,Lindblom:2018rfr,Lindblom:2018ntw} leads to comparable
results, but without the unwanted behavior at the stitching points~\cite{Carney:2018sdv}.

\begin{figure}[]
\includegraphics[width=\columnwidth,clip=true]{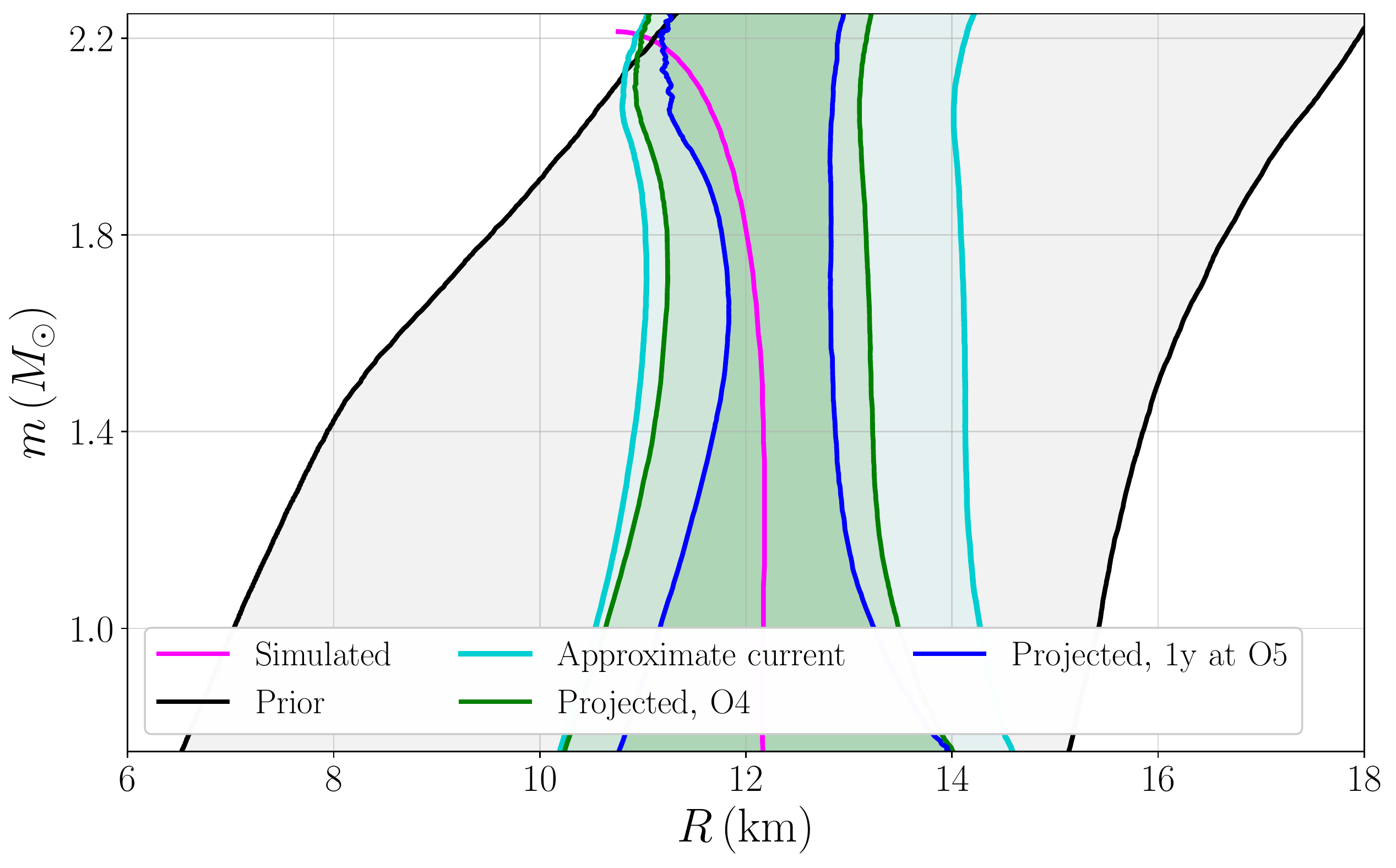}
\caption{Projected equation of state constraints on the mass-radius plane from future expected detections of binary neutron
star coalescences with gravitational waves employing the nonparametric equation of state 
representation of~\cite{Landry:2018prl,Essick:2019ldf}. The pink line shows the simulated equation of state. Black,
cyan, green, and blue lines enclose the 90\% credible interval for the analysis prior, an approximation to the current constraints,
and projected constraints from simulated signal detected during upcoming LIGO/Virgo observing runs respectively.
Adapted from~\cite{Landry:2020vaw}.
}
\label{fig:Rmock}
\end{figure}

Any representation of the equation of state in terms of a finite number of parameters, such as the 
piecewise-polytropic, the spectral, or other parametrizations discussed in Sec.~\ref{EoSpar}, will inevitably lead to loss of 
generality. In contrast, a ``nonparametric" representation, also discussed in Sec.~\ref{EoSpar}, 
based on a Gaussian process conditioned on nuclear models
can reproduce any function, including the true equation of state~\cite{Landry:2018prl,Essick:2019ldf}.
Using this nonparametric approach and an equation of state process conditioned on different sets of nuclear models that
include hadronic, hyperonic, and hybrid hadronic-quark models, Ref.~\cite{Landry:2020vaw} showed that in the next five years
(fifth observing run of LIGO/Virgo circa, 2025~\cite{Aasi:2013wya}) we can obtain ${\cal{O}}(1)$km constraints on neutron star
radii. Figure~\ref{fig:Rmock} presents these projections on a simulated true equation of state (pink line) that is consistent
with current constraints and assuming a realistic set of simulated detected signals. 
The black lines enclose 90\% of the nonparametric prior, while turquoise lines correspond to an approximation
of current constraints when employing information from GW170817, mass-radius measurement of 
J0030+0451~\cite{Miller:2019cac,Riley:2019yda}, and a lower limit on the maximum mass based on heavy pulsars. Green and blue 
lines denote projected constraints from neutron star coalescences potentially observed during the fourth and fifth
 observing~\cite{Aasi:2013wya}, culminating in a ${\cal{O}}(1)$km radius constraint at masses 
 $1.4-1.8M_{\odot}$~\cite{Landry:2020vaw}.

%

%--------------------------------------------------------------------------------------------------
\subsubsection{Effect of the mass distribution and other population parameters}
\label{pop}
 
The hierarchical formalism described in Sec.~\ref{hierarchical} for combining tidal constraints from various gravitational wave signals
(as well as other source of information about neutron star radii and the maximum mass) ensures that priors on the
event-level parameters (for example the tidal deformability of GW170817) do not affect the equation of state constraints.
However, besides the tidal deformability, the inferred neutron star mass might also influence the constraints given the dependence
of tidal effects on the mass, see Fig.~\ref{fig:EoSs}. Since the inferred neutron star mass depends on the assumed mass prior, 
a choice for the prior that does not match the astrophysical neutron star mass distribution could lead to biases in the
equation of state. For example, Ref.~\cite{Agathos:2015uaa} showed large systematic errors in the extracted combined 
$\lambda_{1.4}$ if neutron star masses are around $\sim1.3M_{\odot}$, but analyses assume flat priors over a wider range.

This potential bias shows that prior assumptions about related parameters such as the masses or spins can affect 
equation of state inference. In order to avoid such a bias, one needs a self-consistent hierarchical framework 
that infers the neutron star equation of state together 
with the mass (or spin) distribution of the observed neutron stars in the limit of many
detections~\cite{Wysocki:2020myz,Landry:2020vaw}. 
A misspecified and fixed population model for the neutron star masses could lead to biases in the equation
of state extraction after ${\cal{O}}(25)$ gravitational wave events~\cite{Wysocki:2020myz}. 
While the number of observed gravitational wave signals is so far safely below that number, 
a similar approach that simultaneously models the equation of state and the neutron star mass distribution 
will eventually become essential for X-ray-based radii measurements, or even heavy radio pulsars~\cite{2020arXiv200500032F}.

%--------------------------------------------------------------------------------------------------
\subsubsection{Non gravitational wave messengers}
\label{nicer}

While the main focus of this article is the tidal deformability of neutron stars observed with gravitational wave coalescences, 
future equation of state constraints will undoubtedly include information from other messengers, which we briefly mention here.
The existence of heavy pulsars with masses $\sim2 M_{\odot}$~\cite{Demorest:2010bx,Antoniadis:2013pzd,Cromartie:2019kug}
provides a lower limit on the maximum neutron star mass, which is imposed as a constraint on studies employing an
equation of state representation~\cite{Lackey:2014fwa,Carney:2018sdv,Landry:2020vaw}. Though tempting to incorporate
the lower limit constraint as a sharp cutoff on which equations of state are permissible, Ref.~\cite{Miller:2019nzo} emphasized
the need to marginalize over their measurement uncertainty~\cite{Alvarez-Castillo:2016oln}, as it can be non negligible. 

X-ray pulses from isolated stars observed by NICER can also be used to provide mass and 
radius constraints~\cite{Watts:2016uzu,Miller:2016pom}. Constraints from J0030+0451~\cite{Miller:2019cac,Riley:2019yda}
have already been used in combination with gravitational wave
 measurements~\cite{Miller:2019cac,JiangTang2019,Raaijmakers:2019dks,Landry:2020vaw} to place stronger constraints
 on the neutron star equation of state. Using the NICER target list~\cite{Bogdanov:2019ixe} 
 and current estimates for the rate of binary neutron star detections~\cite{Aasi:2013wya}, Ref.~\cite{Landry:2020vaw} find that NICER
 constraints will be complementary to gravitational waves, especially aiding in constraining soft equations of state.
  The above conclusions can be strengthened by 
 folding in information from nuclear experiment~\cite{Miller:2019nzo} or nuclear calculations~\cite{Forbes:2019xaz}. 
 In particular, Ref.~\cite{Forbes:2019xaz} employs an equation of state parametrization that is 
anchored to nuclear calculations at low densities and transitions to the speed of sound parametrization at higher densities.
 If the equation of state does not exhibit strong first-order phase transitions, then this setup coupled to observations of binary neutron
 coalescences and X-ray based radii measurements by NICER predicts a $\sim20\%$ constraint on the 
 neutron star pressure at densities of $1-2$ times the nuclear saturation in the coming years~\cite{Forbes:2019xaz}.

%--------------------------------------------------------------------------------------------------
\subsubsection{Phase transitions}
\label{PT}

The possibility of a phase transition in neutron star cores towards non nucleonic degrees of freedom poses additional challenges
in the interpretation of tidal deformability constraints. Above a certain high density a 
 phase transition towards deconfined quark matter is expected, though that density might exceed central
neutron star densities, and hence astrophysical neutron stars remain nucleonic~\cite{Oertel:2016bki}. 
Among existing analyses, the ones employing equation of state insensitive relations typically assume the
absense of a strong phase transition~\cite{Zhao:2018nyf,Sieniawska:2018zzj,Carson:2019rjx}.
Some equation of state parametrizations can accommodate phase transitions, including strong ones, for example
the piecewise polytropic one~\cite{Read:2008iy}. The non parametric approach
has the flexibility to use nuclear models with phase transitions as part of the training set, so the resulting synthetic equations 
of state can accommodate this phenomenology~\cite{Essick:2019ldf,Landry:2020vaw}. 
Dedicated parametrizations for phase transitions have also been proposed~\cite{Alford:2013aca,Alford:2015gna,Han:2018mtj}, 
and will be
discussed in Sec.~\ref{EoSpar}.

Overall, a hierarchical inference approach with an appropriate representation of the equation of state can result in
phase-transition-proof results. Going beyond that, the parameter space of phase transitions can be constrained by 
maximum mass~\cite{Chamel:2012ea,Zdunik:2012dj,Alford:2013aca,Alford:2015gna,Han:2018mtj,Han:2020adu}, 
as well as tidal deformability~\cite{Zhang:2019fog} measurements. 
Sufficiently loud detections could be used to identify or rule out such phase transitions from nucleonic matter to, 
for example, kaons, hyperons or quark matter~\cite{Chatziioannou:2015uea}.
Another possibility is that of the potential detection of stars in the ``third family'' (besides neutron stars and white dwarfs)~\cite{Schertler:2000xq}.
 Such stars can have the same 
mass as normal neutron stars from the hadronic family, but differ in radius by a few kilometers. Detection of
such stars could be used to constrain the parameter space of equations of state admitting strong phase transitions~\cite{Alvarez-Castillo:2016oln}.

The detection of a population of coalescing binary neutron star signals can put the observed ``universality" of 
hadronic equations of state, Fig.~\ref{fig:universality}, to the test
~\cite{Chen:2019rja,Chatziioannou:2019yko}.  A deviation from
the almost-constant-radius assumption or the expected chirp mass-chirp radius behavior could act as a clear 
indication of a transition away from a hadronic equation of state.
If this transition happens in the density/mass regime of neutron star coalescences, 
 the transition mass could be estimated with $\sim 100$ detections~\cite{Chatziioannou:2019yko}. If instead, as is 
 possible, the transition happens at sufficiently high densities such as that neutron stars are minimally affected, then gravitational wave
 observations can place limits on the relevant parameter space of the transition onset and strength~\cite{Chatziioannou:2019yko}.

%--------------------------------------------------------------------------------------------------
\subsubsection{Neutron star-black hole coalescences}
\label{sec:nsbh}

Mixed binary systems of a neutron star and a black hole have not been unambiguously detected yet 
with either gravitational or electromagnetic radiation, however, the emitted gravitational wave signal during the late
stages of such a coalescence might still carry the signature of tidal effects~\cite{Lackey:2013axa,Foucart:2013psa}. 
The neutron star binary component will be tidally distorted by the field of the black hole companion, if the tidal field is strong enough
to deform the neutron star before it plunges into its companion. Favorable conditions for tidal disruption include a small
black hole (such that it possesses a large tidal field), with a large spin (so that the neutron star can orbit closer to the black
hole before plunging), and a less compact neutron star (such that it is more
 deformable)~\cite{Lackey:2013axa,Foucart:2013psa,Foucart:2018lhe,Foucart:2018rjc}. 

The prospects for equation of state constraints from neutron star-black hole coalescences depends on the degree to which
the neutron star is disrupted.
If the neutron star is not significantly distorted before plunging into the black hole, the resulting signal will only carry  
minimal imprints from finite-size effects, suggesting that it might be difficult to unambiguously 
establish the nature of such a system observationally. If, on the other hand, the neutron star
experiences significant disruption, then the leading-order tidal effect in the gravitational phase 
is again given by Eq.~\ref{lambdaTdef} for $\tilde{\Lambda}$, but where now the the tidal deformability 
of one of the binary components -the black hole- is zero. At the same time, the amplitude of the emitted signal
carries a stronger matter imprint than in the case of neutron star coalescences,
as now a total neutron star disruption will effectively serve to prematurely terminate the
 signal~\cite{Lackey:2013axa,Pannarale:2013uoa,Pannarale:2015jka}. 
 
\begin{figure}[]
\includegraphics[width=\columnwidth,clip=true]{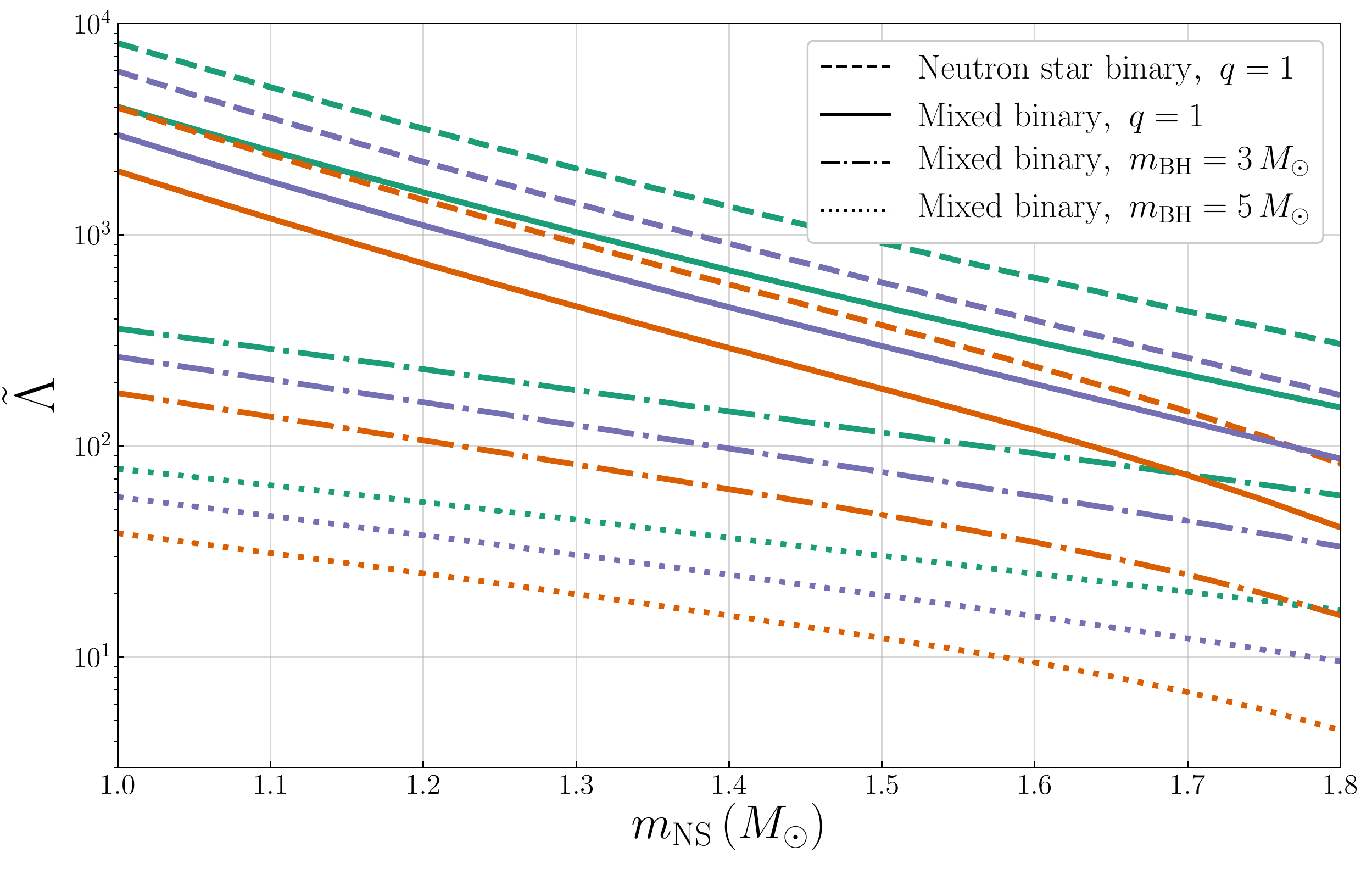}
\caption{Dependence of $\tilde{\Lambda}$ on the type of binary as a function of the mass of the neutron star for different random
equations of state (different colors). Dot-dashed and dotted lines correspond to different black hole masses, while the
dashed and solid lines show $\tilde{\Lambda}$ for an equal-mass neutron star and mixed binary respectively. 
Tidal effects are stronger in neutron star-black
hole systems when both the neutron star and the black hole are light.
}
\label{fig:LTnsbh}
\end{figure}

 For systems undergoing tidal disruption, both the frequency of disruption~\cite{Pannarale:2015jia} and tidal effects in the 
 phase~\cite{Kumar:2016zlj} could be used to place constraints on the neutron star equation of state. 
 Compared to neutron star binaries, the gravitational wave phase evolution of mixed neutron star-black hole binaries 
 is less affected by tidal effects as $\tilde{\Lambda}$ is smaller due to both the typically asymmetric mass ratio and 
 $\lambda_{\mathrm{BH}}=0$. Figure~\ref{fig:LTnsbh} shows the dependence of 
 $\tilde{\Lambda}$ on the type of binary and its mass ratio.
 For a given equation of state, neutron star binaries result in a larger value of $\tilde{\Lambda}$ (dashed lines) than mixed
 binaries of the same mass (solid lines) as the latter have $\lambda_{\mathrm{BH}}=0$. For more astrophysically plausible
 black hole masses, the asymmetric mass ratio leads to further reduction of $\tilde{\Lambda}$ (dot-dashed and dotted lines).
 The most favorable systems for measuring $\tilde{\Lambda}$ with mixed binaries are consisted of the least massive neutron 
 stars and black holes. 
 Observing 20-35
  disruptive systems could lead to a $25-50\%$ measurement of the neutron star tidal deformability, though the exact
 constraints depend on the mass distribution of such systems~\cite{Kumar:2016zlj}.
 In parallel, the rich phenomenology and possible outcomes of neutron star-black hole coalescences result in a wide range of
 possible electromagnetic counterparts that can provide complementary constraints on neutron 
 star matter~\cite{Ascenzi:2018mwp,Barbieri:2019sjc,Barbieri:2019kli,Barbieri:2019bdq}.
 Accurate waveform models for such systems must also account for their possible outcomes, 
 and ongoing progress~\cite{Thompson:2020nei,Matas:2020wab} is essential in order to keep systematic errors under
 control~\cite{Chakravarti:2018uyi,Huang:2020pba}.

The fact that only a single tidal parameter can be extracted from neutron star coalescences~\cite{Wade:2014vqa} and uncertainties
about the maximum possible neutron star mass suggests that it is difficult to unambiguously distinguish between 
neutron star-black hole and binary neutron star coalescences~\cite{Yang:2017gfb}.  
If one further entertains the possibility of
$\sim1.35M_{\odot}$ black holes, both the gravitational and electromagnetic signal from GW170817 are consistent with 
a neutron star-black hole coalescence~\cite{Hinderer:2018pei,Coughlin:2019kqf} and similar 
for GW190425~\cite{Han:2020qmn,Kyutoku:2020xka}. This possibility has led to increased interest in equal-mass neutron star-black
hole coalescences and their potential counterparts~\cite{Foucart:2019bxj}.  
For such low mass systems, though, tidal effects in the gravitational signals from dozens of detections could lead to 
constraints on the abundance of a potential low mass neutron star-black hole population~\cite{Chen:2019aiw}.
 Additionally, more massive neutron stars experience inherently weaker tidal effects, see Fig.~\ref{fig:EoSs}, which are 
 harder to measure. For systems such as GW190425 with a total mass of $M \sim 3.4M_{\odot}$, we might not be 
 able to establish the presence of tidal 
effects in the signal until next-generation detectors~\cite{Chen:2020fzm}

%%%%%%%%%%%%-------------------------------------------------
\subsection{Generic equation of state representations}
\label{EoSpar}

Before proceeding to GW170817 and GW190425, here we briefly discuss various generic equation of state representations
that are used to analyze astronomical data.
Despite the wide range of nuclear equation of state models of different physical content that have been proposed throughout the years, 
all models need to share some characteristics. These characteristics include causality 
(the speed of sound must be smaller than the speed of light throughout the
star), thermodynamic stability (the pressure must be a monotonically increasing 
function of the energy density), and consistency with
known heavy pulsars (the equation of state must support stable neutron stars of 
$\sim 2M_{\odot}$~\cite{Demorest:2010bx,Antoniadis:2013pzd,Cromartie:2019kug}).

Given these common characteristics and in order to systematize the study of equations of state, a number of parametrizations 
have been proposed that express the pressure as a function of the density in terms of a few tunable parameters. 
These parametrizations are not specific to gravitational wave analyses, and they have been employed in the 
interpretation of X-ray data as well~\cite{Steiner:2010fz,Miller:2019cac,Raaijmakers:2019qny,Morawski:2020izm}.
The number of tunable parameters
needs to be large enough such that the parametrization can represent target (or even the true) equations of state well-enough 
(for an appropriate definition of ``well-enough"). At the same time, the number of parameters needs to be low enough in order for 
the parametrization to be meaningfully constrained from the observations we realistically expect to obtain. In other words,
a parametrization needs to fit nuclear models, but not overfit them.

Perhaps the simplest parametrization expresses the equation of state in terms
of a series of polytropes of the form $p\sim \rho^{\Gamma}$ where $p$ is the pressure, $\rho$ is the mass density, and $\Gamma$ is the adiabatic index.
Read et al~\cite{Read:2008iy} argued that three polytropes
with fixed transition densities strike a good balance between accuracy and parsimony.
The transition densities are selected by optimizing the fit to target equation of state models, resulting in a piecewise 
polytropic parametrization. Given the nature of the polytropes, 
this parametrization can reasonably reproduce  equations of state with phase transitions for an 
appropriate choice of the adiabatic indices. Generalizations to this form of the piecewise polytropic parametrization
include replacing the lowest density polytrope with a series expansion of the symmetry 
energy around saturation density~\cite{Steiner:2010fz}, extending to five polytropic segments~\cite{Raithel:2016bux,Raithel:2017ity},
parametrizing the neutron star crust~\cite{Read:2008iy},
and ensuring a smooth speed of sound~\cite{OBoyle:2020qvf}.

In contrast to the above parametrizations that employ polytropes to some extent, a 
``spectral parametrization"~\cite{Lindblom:2010bb,2012PhRvD..86h4003L,Lindblom:2013kra,Lindblom:2018rfr,Lindblom:2018ntw}
expresses the adiabatic index $\Gamma$ of the equation of state (rather than the pressure itself)
in terms of free parameters. One advantage of the spectral parametrization is that it can automatically accommodate for
stability~\cite{Lindblom:2010bb} and causality~\cite{Lindblom:2018rfr}, and result in smooth models which present an advantage from a 
data analysis perspective~\cite{Carney:2018sdv}. The spectral parametrization produces more accurate fits for smooth
equations of state than piecewise polytropes, though it has been shown to be able to model phase 
transitions to some extent as well~\cite{Lindblom:2010bb}.

The issue of phase transitions and the physical properties of neutron star matter have also given rise to the constant-sound-speed 
parametrization~\cite{Alford:2013aca,Alford:2015gna,Han:2018mtj}, 
and the speed of sound parametrization~\cite{Tews:2018kmu}. The former is tailored to phase transitions towards
deconfined quark matter and can be 
incorporated on top of a primarily hadronic parametrization such as the spectral one. It introduces three additional parameters in
order to capture the phenomenology of first-order phase transitions: the energy density at which the transition occurs $\epsilon_{trans}$,
the strength of the transition $\Delta \epsilon$, and the constant speed of sound in the quark phase. The speed-of-sound
parametrization, instead, expresses the speed of sound $c_s$ in terms of parametric functions selected in order to reproduce
desirable phenomenology, see Fig. 1 of~\cite{Tews:2018kmu}. This ensures that the speed of sound asymptotes to the 
conformal limit of $c_s^2=1/3$ in the high density regime, but can possibly violate it for intermediate densities.

A different approach to construct synthetic equation of state models without being limited to a certain parametrized family
is based on Gaussian processes~\cite{Essick:2019ldf,Landry:2020vaw}.
Parametric equation of state representations only have support for $p(\epsilon)$ functions within their range,
assigning zero prior probability on any other function. The true equation of state need not adhere to any of
our parametrizations, so it can only be reproduced approximately and to within some systematic uncertainty by them.
An extension of parametrized equations of state that instead assigns nonzero prior probability on any possible function
does not rely on a specific parametrization, but instead employs a Gaussian process conditioned on nuclear 
models~\cite{Landry:2018prl,Essick:2019ldf}. This nonparametric approach is based on a number of
hyperparameters that control the function correlations (for example, the correlations between pressures
at different densities) that are marginalized over~\cite{Essick:2019ldf}. 
The result is synthetic causal and thermodynamically stable equations of state that resemble the input nuclear models to a
tunable degree. Depending on the input models, the synthetic equations of state can account for non-hadronic degrees of
freedom or nuclear calculations~\cite{Essick:2020flb}.

As also emphasized in Sec.~\ref{hierarchical} on hierarchical analyses, 
when the inference is formulated in terms of an equation of state representation (parametric or not), 
the prior distribution is specified on the equation of state directly, rather than the macroscopic properties of the
observed systems. Due to the complicated relation between the equation of state representations and the various
quantities of interest, such as neutron star radii and maximum mass, the resulting prior on quantities of interest might be non trivial or 
even exhibit undesirable features, such as being overly
informative in untested regions of the parameter space. This was exemplified in~\cite{Greif:2018njt} where prior choices in
two parametrizations (the speed of sound parametrization and
 the piecewise polytropes) were unsurprisingly shown to affect the resulting inference. 
 It is therefore important to be cognizant of the equation of state priors used
in each analysis and how they propagate to derived quantities, especially when comparing results from different studies.

%%%%%%%%%%%%-------------------------------------------------
\subsection{Hierarchical inference}
\label{hierarchical}

As a final detour before delving into GW170817 and GW190425, here we briefly introduce the framework of hierarchical inference
for combining information from different sources. The framework for combining information from different source including their 
statistical errors and selection effects has a long history~\cite{Loredo:2004nn,Bovy:2009kp,Hogg:2010ma,2014ApJ...795...64F} and has already been used successfully in the gravitational 
wave context to measure the population properties of merging 
black holes~\cite{Mandel:2009pc,TheLIGOScientific:2016htt,Stevenson:2017dlk,LIGOScientific:2018jsj}. As the name suggests, 
analyses happen on different levels: at the bottom level we have the raw gravitational wave data, which are used to constrain the
binary system properties on the next level. On the final level, event-specific constraints from multiple signals are used to constrain
some population property of the systems, such as the common equation of state or the astrophysical mass distribution.

The subsequent discussion follows~\cite{Landry:2020vaw}, however, similar discussion has also been 
presented in~\cite{Lackey:2014fwa,Miller:2019nzo,Wysocki:2020myz}. Though we specialize to gravitational waves, a similar
approach can be employed for different types of astronomical data, such as NICER, radio pulsars, X-ray binaries, etc., 
see Sec IIIb of~\cite{Landry:2020vaw}. The goal is to compute the posterior probability for an 
equation of state $\epsilon$ (constructed either parametrically or nonparametrically) given some gravitational wave data 
$d=\{d_1...d_i\}$ for multiple relevant $i$ detections:
\begin{equation}
p(\epsilon|d)=\frac{p(\epsilon)p(d|\epsilon)}{p(d)},\label{BT}
\end{equation}
where $p(\epsilon|d)$ is the desired posterior, $p(\epsilon)$ is the prior on the equation of state, 
$p(d|\epsilon)$ is the likelihood whose form depends on the type of data under consideration, and $p(d)$ the evidence.
For independent observations (for example GW170817 and GW190525), the likelihood can be factorized as
$p(d|\epsilon)=\prod_i p(d_i|\epsilon)$, where the likelihood of each event given the equation of state is
\begin{align}
p(d_i|\epsilon) &= \int dm_{1i} \,dm_{2i} \int \Lambda_{1i} \,\Lambda_{2i} \int d\theta \,p(m_{1i},m_{2i},\Lambda_{1i} \Lambda_{2i}|\epsilon,\theta)\nn\\
&\times \frac{p(d_i|m_{1i},m_{2i},\Lambda_{1i} \Lambda_{2i})}{\beta(\theta)},
\end{align}
where $m_{1i},m_{2i}$ are the binary component masses, $\Lambda_{1i},\Lambda_{2i}$ are the binary component dimensionless
tidal deformabilities, and $\theta$ collectively denote population parameters for neutron star binaries, for example the slope
of the neutron star mass function or the maximum neutron star spin. The posterior for the equation of state is obtained by marginalizing
over these population parameters, but they still affect the overall inference. 
As already discussed, neglecting to marginalize over the neutron star mass population parameters
and assuming they are fixed can lead to biases~\cite{Wysocki:2020myz}. The term $\beta(\theta)$ describes the selection
effects of gravitational wave observatories, primarily the preference for more massive binaries that emit 
stronger signals~\cite{Mandel:2018mve}. The term $p(d_i|m_{1i},m_{2i},\Lambda_{1i} \Lambda_{2i})$ is the usual likelihood
for the mass and tidal parameters of the observed binary. 

The term $p(m_{1i},m_{2i},\Lambda_{1i} \Lambda_{2i}|\epsilon,\theta)$ links the masses and tidal parameters of the binary to 
the equation of state $\epsilon$. Given that the equation of state unambiguously defines the tidal deformability of a neutron star of 
a given mass, this term simplifies to
\begin{equation}
p(m_{1i},m_{2i},\Lambda_{1i} \Lambda_{2i}|\epsilon,\theta) = p(m_{1i},m_{2i}|\epsilon,\theta) \delta(\Lambda_{1i}-\Lambda(\epsilon,m_{1i}))
 \delta(\Lambda_{2i}-\Lambda(\epsilon,m_{2i})),
\end{equation}
making the integral over $\Lambda_{1i},\Lambda_{2i}$ trivial:
\begin{align}
p(d_i|\epsilon) &= \int dm_{1i} \,dm_{2i} \int d\theta \,p(m_{1i},m_{2i}|\epsilon,\theta) \frac{p(d_i|m_{1i},m_{2i},\Lambda(\epsilon,m_{1i}) \Lambda(\epsilon,m_{2i}))}{\beta(\theta)}\label{hierLfull}.
\end{align}
If we further want to restrict to a fixed population $\theta$, an acceptable approximation for inference from 
$\lesssim20$ sources~\cite{Wysocki:2020myz}, the likelihood simplifies to
\begin{align}
p(d_i|\epsilon) &\sim \int dm_{1i} \,dm_{2i}\,p(m_{1i},m_{2i}|\epsilon)\, p(d_i|m_{1i},m_{2i},\Lambda(\epsilon,m_{1i}) \Lambda(\epsilon,m_{2i})).\label{hierL}
\end{align}
Equations~\ref{hierL} and~\ref{BT} can be directly used to infer the underlying equation of state from a list of binary
neutron star observations. The first term in the integral $p(m_{1i},m_{2i}|\epsilon)$ is a prior distribution over the system masses
and spells out why the resulting inference will depend on the binaries' population parameters. In general, this dependence needs
to be modeled and marginalized over through Eq.~\ref{hierLfull}. The second term $p(d_i|m_{1i},m_{2i},\Lambda(\epsilon,m_{1i}) \Lambda(\epsilon,m_{2i}))$ expresses the information gained from the binary neutron star observation. In practice, analyses
of neutron star binaries compute and publicly release samples from the posterior distribution of these
parameters $p(m_{1i},m_{2i},\Lambda(\epsilon,m_{1i}) \Lambda(\epsilon,m_{2i})|d_i)$~\cite{170817samples,190425samples}. 
In order to obtain the likelihood from them,
the custom prior on the neutron star masses and tidal deformabilities that each analysis used must be taken into account 
and removed.

Notably absent from the above framework and Eq.~\ref{hierL} is any prior on the tidal deformabilities of 
the coalescing neutron stars observed. The only relevant priors entering the above equations are the ones 
on the equation of state in Eq.~\ref{BT} 
and the dependence on the neutron star population properties (which can and should eventually be marginalized over). This is an
elegant outcome of hierarchical inference as the event-level priors on the parameters of interest, for example the tidal
deformability of merging neutron stars or the radii of NICER targets, do not affect the next-level constraints, namely the equation
of state.

%%%%%%%%%%%%%%%%%%%%%%%%%%%%%%%
\section{The binary neutron star coalescence signals GW170817 and GW190425}
\label{170817}

GW170817~\cite{TheLIGOScientific:2017qsa} and GW190425~\cite{Abbott:2020uma} are to date the only detected 
gravitational wave signals consistent with the coalescence of two neutron stars. The $\sim 2$ minute-long duration of the signals
in the detectors' sensitive band immediately suggests that the binary components are less massive than most other
 detections~\cite{2018arXiv181112907T}. Estimated masses lie in the $1-2M_{\odot}$ range, making them
 consistent with known
 galactic neutron star masses~\cite{Tauris:2017omb}.
 Definitive proof of the nature of the binaries with gravitational waves alone would
 require the detection of tidal effects from two bodies, something challenging with current-sensitivity
  detectors~\cite{DelPozzo:2013ala,Wade:2014vqa,Yang:2017gfb,LIGOScientific:2019eut}. 
  The electromagnetic counterpart to GW170817 guarantees the 
  presence of at least one neutron star in the system, but the possibility of a neutron star-black hole binary cannot be 
  ruled out~\cite{Hinderer:2018pei,Coughlin:2019kqf}. The GW190425 source is consistent with containing one or even two
  low-mass black holes.

A detector glitch in the LIGO-Livingston detector overlapped with the GW170817 signal shortly before merger and at around
 $100-200$Hz~\cite{TheLIGOScientific:2017qsa}.
The overlap frequency is lower than the typical signal frequencies where tidal effects become important, see Fig.~\ref{fig:waveform},
however, it could potentially influence the measurement of other binary parameters. 
Eventually, the instrument glitch was coherently fit
and subtracted from the LIGO-Livingston data, 
allowing us to regain access to the underlying signal~\cite{Cornish:2014kda,Littenberg:2014oda}. 
A further study involving
simulated signals from neutron star binaries placed on top of other detector glitches showed 
that the subtraction procedure is robust and does not bias
estimates of binary properties~\cite{Pankow:2018qpo}.

The chirp mass of the two binaries was measured to be 
 $1.186^{+0.001}_{-0.001}M_{\odot}$ and $1.44^{+0.02}_{-0.02}M_{\odot}$ for GW170817 and GW190425 respectively 
 at the 90\% level, while both
 signals are consistent with equal component masses, no spin, and no relativistic spin-precession. Estimates on the total mass
 of the binaries depend on assumptions about the component spins as the two are correlated~\cite{PhysRevD.49.2658}.
Assuming small spins $\chi<0.05$ inspired by galactic observations~\cite{Tauris:2017omb} results in estimates of the total mass
of $2.73^{+0.04}_{-0.01}M_{\odot}$ and $3.3^{+0.1}_{-0.1}M_{\odot}$ respectively, suggesting that GW170817 is consistent
with known neutron star binaries in the Galaxy, while GW190425 is a $\sim 5\sigma$ outlier~\cite{Abbott:2020uma}.
Allowing for high spins $\chi<0.89$ naturally increases the measurement uncertainty in the total masses to 
$2.77^{+0.22}_{-0.05}M_{\odot}$ and $3.4^{+0.3}_{-0.1}M_{\odot}$ respectively.

%%%%%%%%%%%%-------------------------------------------------
\subsection{Tidal properties: generic analysis}

The first step in understanding systems such as GW170817 and GW190425 involves a generic analysis that imposes no constraints
on the relationship between the tidal deformabilities of the coalescing objects $\Lambda_1$ and $\Lambda_2$ and as a consequence, no restrictions 
on their nature. In practice, $\Lambda_1$ and $\Lambda_2$ are assumed to be independent of each other, with no prior restrictions
on the relation between them, allowing the coalescing bodies to be black holes, exotic compact objects, or in general to not
follow a relation that resembles the equations of state in Fig.~\ref{fig:EoSs} in any way. This analysis results
in a constraint on $\tilde{\Lambda}$ -the only measured tidal parameter- of $\tilde{\Lambda} \lesssim 700$ at the 90\% for
 GW170817~\cite{TheLIGOScientific:2017qsa,Abbott:2018wiz} and similar for GW190425~\cite{Abbott:2020uma}.
Further studies have analyzed the detector data in this setup assuming independent tidal components and have 
obtained consistent results~\cite{Dai:2018dca,Narikawa:2018yzt,Narikawa:2019xng,Han:2020jvr}.

Despite resulting in comparable upper limits on $\tilde{\Lambda}$, GW190425 did not lead to novel constraints of
neutron star matter due to its high mass. Figure 14 in~\cite{Abbott:2020uma} compares the GW190425 $\tilde{\Lambda}$
constraint from the data to the expected value for a system of GW190425's mass given the GW170817 equation of state 
constraints, suggesting that the upper limit is not competitive. Despite this, GW190425 is important for equation of state studies
as it demonstrates that heavy neutron stars form binaries and merge and can thus be detected with gravitational waves.
A signal similar to GW190425 but stronger could lead to constraints on the high density equation of state.

\begin{figure}[]
\includegraphics[width=\columnwidth,clip=true]{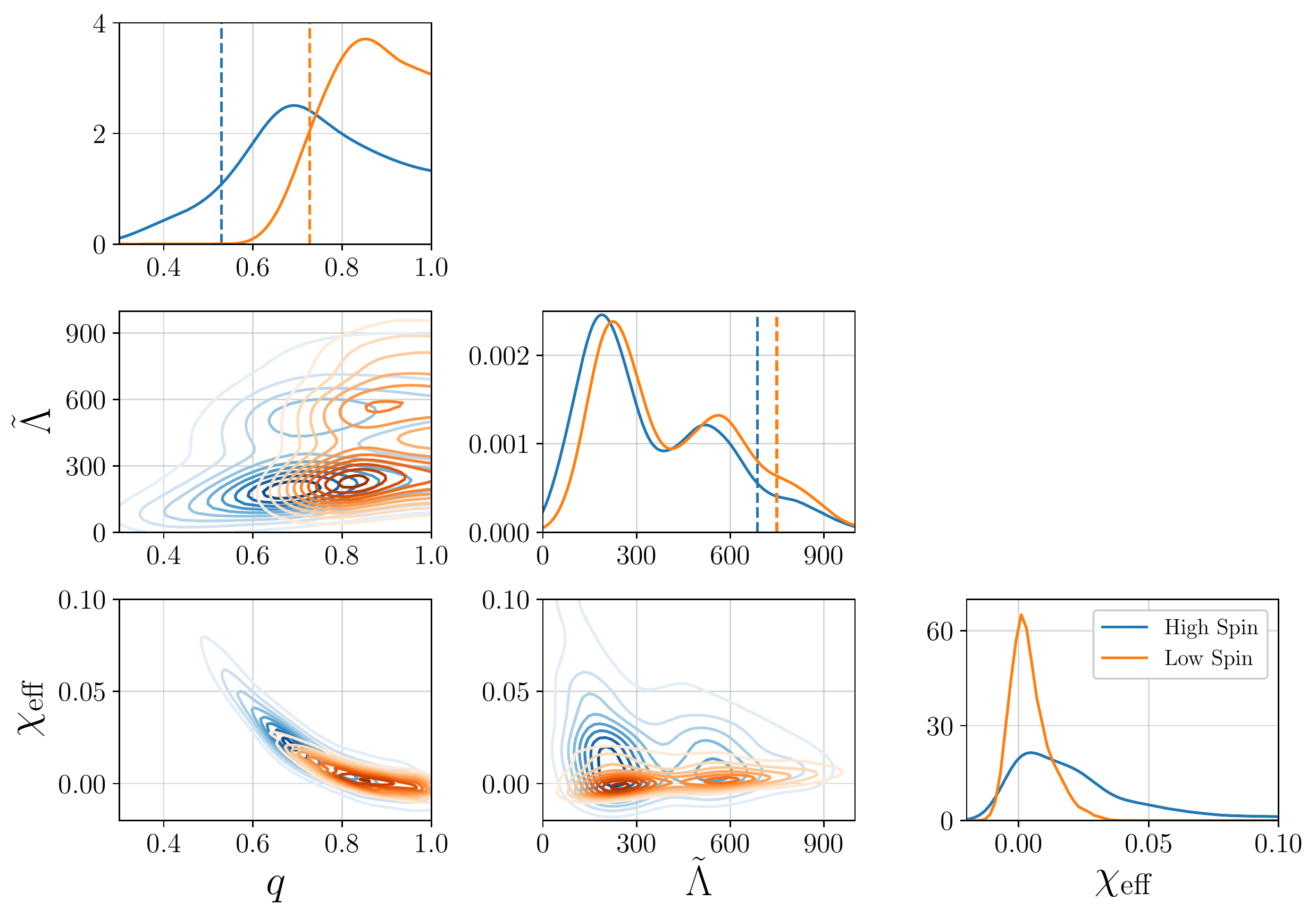}
\caption{Marginalized 1- and 2-dimensional posterior distributions for the mass ratio $q$, the effective spin parameter $\chi_\mathrm{eff}$, and the tidal
parameter $\tilde{\Lambda}$ for GW170817 under the high (blue) and low (orange) spin prior. Dashed vertical lines denote 90\% upper or lower limits as applicable.
Posterior samples correspond to the analysis of~\cite{Abbott:2018wiz} and are available from~\cite{170817samples}.
}
\label{fig:170817real}
\end{figure}

Figure~\ref{fig:170817real} shows the marginalized 2- and 1-dimensional posterior distributions on the mass ratio $q$, 
the effective spin parameter $\chi_\mathrm{eff}$, and the tidal parameter $\tilde{\Lambda}$ for GW170817. The effective 
spin parameter is a specific spin combination that is conserved under spin-precession to at least 2PN order~\cite{Racine:2008qv}.
The bottom left panel shows the characteristic mass-spin correlation that deteriorates measurement of both, while 
$\tilde{\Lambda}$ is less correlated with either $q$ or $\chi_\mathrm{eff}$. These results are obtained with the 
{\tt IMRPhenomPv2\_NRTidal}~\cite{Dietrich:2018uni} waveform model, the only model available 
that includes both tidal and spin-precession
effects, and whose tidal sector has been fitted to numerical simulations~\cite{Dietrich:2017aum}. 
Results obtained with different models that differ in their point-particle sector, spin interactions, and tidal effects
suggest that waveform systematic errors are subdominant compared to statistical errors for GW170817 and GW190425, 
and constraints are robust~\cite{Abbott:2018wiz,2018arXiv181112907T,Narikawa:2019xng,Abbott:2020uma}. 
See~\cite{Dietrich:2020eud} for a review of different waveform models and their relations.

%%%%%%%%%%%%-------------------------------------------------
\subsection{Tidal properties: constrained analysis assuming a neutron star binary}

\begin{figure}[]
\includegraphics[width=\columnwidth,clip=true]{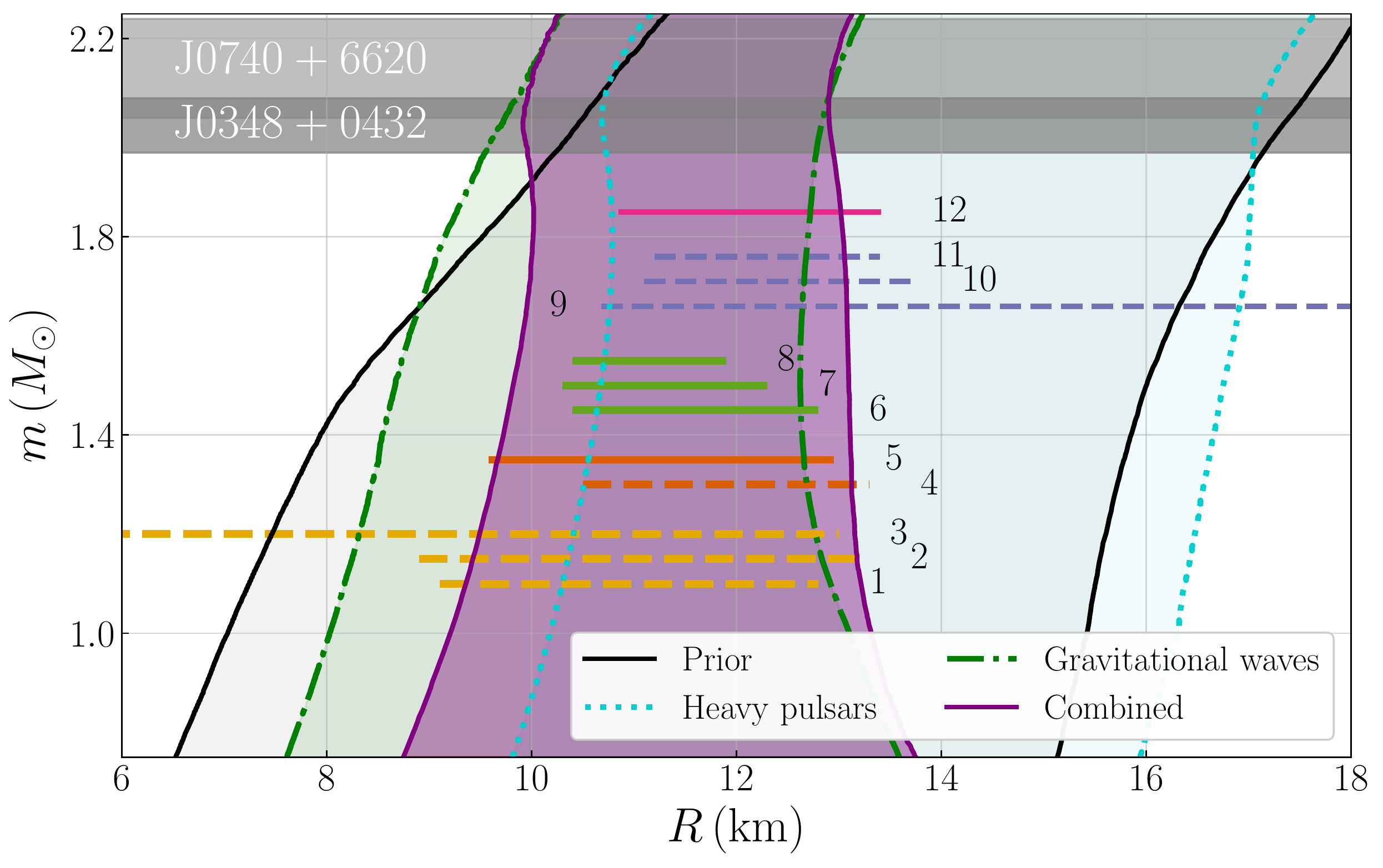}
\includegraphics[width=\columnwidth,clip=true]{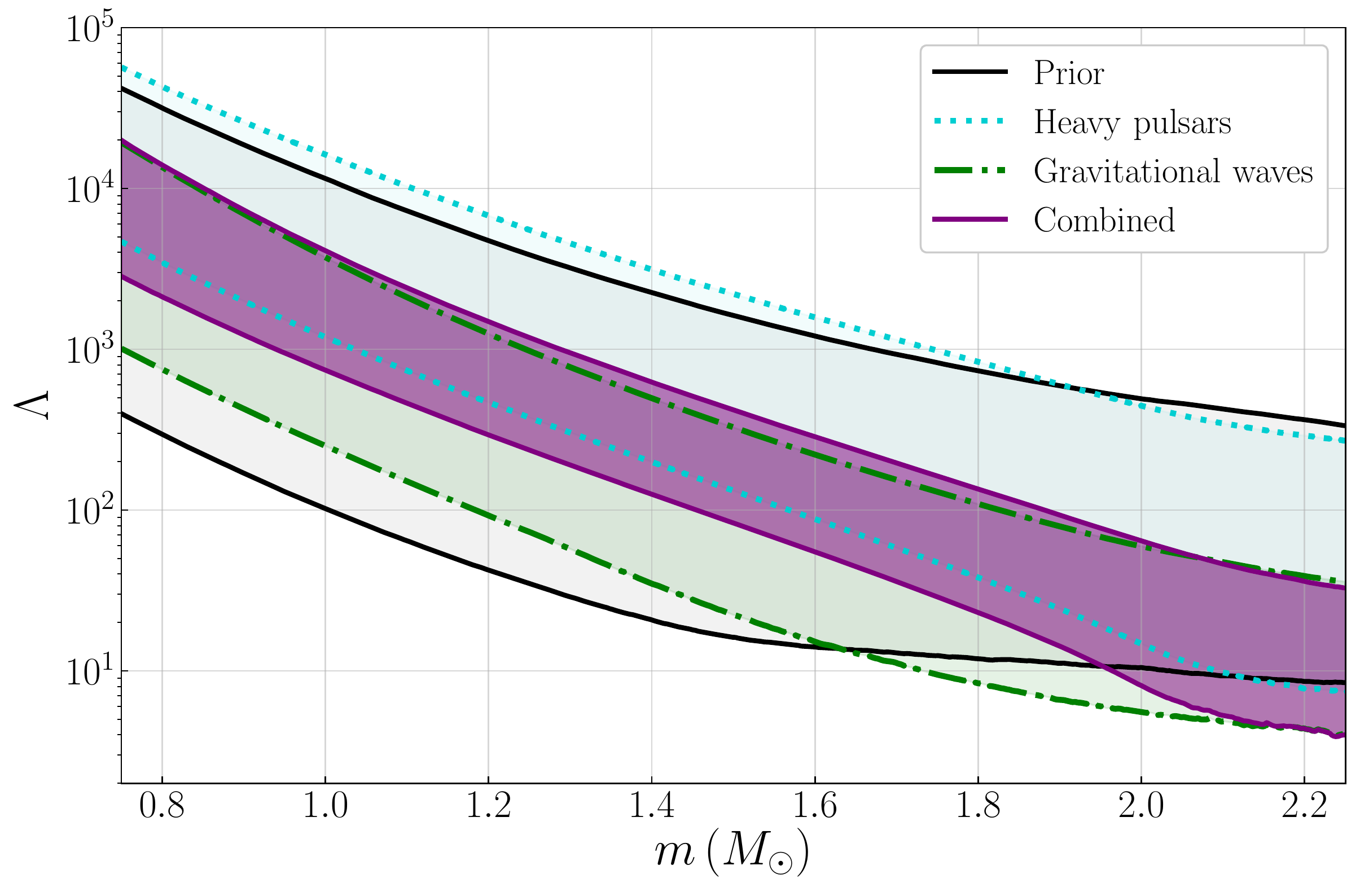}
\caption{Mass-radius (top) and tidal deformability-mass (bottom) constraints on the equation of state. Black,
cyan, green, and purple lines and shaded regions enclose the 90\% credible interval of the prior, and the posterior after
incorporating information from the heavy pulsars, the gravitational wave signals, and a combination of the two respectively
using the nonparametric equation of state representation and hierarchical analysis from~\cite{Landry:2020vaw}.
The grey shaded bands show the 
1-$\sigma$ mass measurement for J0348+0432~\cite{Antoniadis:2013pzd} and J0740+6620~\cite{Cromartie:2019kug}. 
Horizontal lines on the top panel (arbitrary ordinate) denote radius constraints obtained
with different methodologies and assumptions. Solid (dashed) lines correspond to constraints that are (are not) applicable for 
equations of state with strong phase transitions. See the text for details and discussion. Yellow lines show constraints using equation of state insensitive relations: 1~\cite{Abbott:2018exr}, 2~\cite{De:2018uhw}, 3~\cite{Raithel:2018ncd}. Orange lines show constraints using 
equation of state representations: 4~\cite{Abbott:2018exr}, 5~\cite{Landry:2020vaw}. Green lines show constraints that incorporate
chiral effective field theory calculations: 6~\cite{Essick:2020flb}, 7~\cite{Capano:2019eae} (once saturation), 8~\cite{Capano:2019eae} (twice 
saturation). Blue lines show multimessenger constraints: 9~\cite{Bauswein:2017vtn}, 10~\cite{Radice:2018ozg}, 11~\cite{Coughlin:2018fis}.
Pink lines show constraints that incorporate the recent NICER results: 12~\cite{Landry:2020vaw}.
}
\label{fig:radius}
\end{figure}

Going beyond these first results, assuming that GW170817 is a neutron star coalescence can result in more stringent 
constraints on its properties by eliminating part of the parameter space that might be consistent with the data, but not 
consistent with a neutron star coalescence. In the following discussion of such studies
it is worth keeping in mind that all analyses use the same data (modulo
a recalibration between~\cite{TheLIGOScientific:2017qsa} and~\cite{Abbott:2018wiz}), namely the 
gravitational wave signal for GW170817\footnote{Studies that reanalyze the gravitational wave data directly
~\cite{Dai:2018dca,Narikawa:2018yzt,De:2018uhw,Narikawa:2019xng,Han:2020jvr} rather than 
interpret the original constraints might also have differences 
in the waveform and noise model employed, the analysis bandwidth, and choice of stochastic sampler. All these should 
be subdominant compared to statistical errors.}.
Therefore any differences between the results are solely due to the different
prior each analysis employs to either link the component tidal deformabilities or to restrict to specific equation of state
models~\cite{Abbott:2018exr,Kastaun:2019bxo}. Figure~\ref{fig:radius} summarizes these results.
 
The equation of state insensitive relations between macroscopic neutron star properties discussed in Sec.~\ref{LNS} can be used to obtain 
constraints from GW170817 given the assumptions of these relations, namely that the binary is consisted of two neutron stars that share the same
hadronic equation of state. The expectation that $\Lambda(m)\sim m^{-5}$~\cite{DelPozzo:2013ala} for hadronic equations of state, 
see also Fig.~\ref{fig:EoSs}, can be used to translate the generic $\tilde{\Lambda}$ constraint into a constraint on $\Lambda_{1.4}<800$~\cite{TheLIGOScientific:2017qsa} by means of a linear expansion of $\Lambda(m)m^5$ around $m=1.4M_{\odot}$.
The approximate independence of $\tilde{\Lambda}$ on the binary mass ratio given the extremely well-measured binary chirp
 mass for hadronic equations of state~\cite{Wade:2014vqa}, in turn, suggests that the $\tilde{\Lambda}<800$ constraint~\cite{TheLIGOScientific:2017qsa} 
 translates to
  $R<13$km~\cite{Raithel:2018ncd}. 
  
 Besides post-processing results from the generic analyses of~\cite{TheLIGOScientific:2017qsa,Abbott:2018wiz}, equation of state insensitive
 relations can be imposed during stochastic sampling from the system parameter posterior distribution. Along those lines, assuming that
 $\Lambda(m)\sim m^{-6}$~\cite{Zhao:2018nyf} and the radius of neutron stars is approximately constant in the relevant mass range leads to
 $8.9 <R/\mathrm{km} < 13.2 $ at the 90\% level~\cite{De:2018uhw}. At the same time,~\cite{De:2018uhw} established that assuming a prior on
 the component masses inspired by galactic observations has a minimal effect on the resulting radius constraints. Instead, imposing the relation
 of~\cite{Yagi:2015pkc,Chatziioannou:2018vzf} between the component dimensionless tidal deformabilities and the mass ratio of the binary,
 the radius of the most massive component is $R_1=10.8^{+2.0}_{-1.7}$km at the 90\% level and similar for the least massive neutron star
 ~\cite{Abbott:2018exr}. This analysis also updated the fiducial tidal deformability estimate to $\Lambda_{1.4}=190^{+390}_{-120}$.
Both the above results are valid under the assumption that GW170817 originated from a neutron star binary and the equation of state does not 
exhibit a strong phase transition in the relevant mass/density range.

In order to translate from these macroscopic constraints to constraints on the microscopic equation of state, a specific representation of the
equation of state needs to be employed, see Sec.~\ref{EoSpar}. Analyses along those lines still assume that both coalescing bodies are neutron
stars, while the possibility of strong phase transitions depends on the parametrization. The spectral parametrization was employed in~\cite{Abbott:2018exr} (the results were shown to be consistent with piecewise polytropes as well) to reanalyze the strain data, 
and lead to an estimate of the radius that also takes into account that the equation of state 
must support at least $1.97M_{\odot}$ neutron stars~\cite{Antoniadis:2013pzd}. The combined result is $R_1\sim R_2 = 11.9^{+1.4}_{-1.4}$km
at the 90\% level, suggesting that the existence of heavy pulsars narrows down neutron star radii estimates by about $1$km on the low side. 
The neutron star pressure at twice saturation was constrained to $3.5^{+2.7}_{-1.7} \times10^{34}$dyn/cm$^2$, which was shown to be 
consistent with low density results from terrestrial heavy-ion collision experiments~\cite{Danielewicz:2002pu,Russotto:2016ucm}
in~\cite{Tsang:2018kqj}.
The treatment of the neutron star crust~\cite{Fortin:2016hny} in the polytropic and spectral representations affects 
$\Lambda$ to less than 1\%~\cite{Biswas:2019ifs,Kalaitzis:2019dqc}, and similarly for the radius.
For GW170817 and using the spectral representation, the crust model affects the radius
to $100-200$m~\cite{Gamba:2019kwu}, which is smaller than the statistical error of $2-3$km at the 90\% level.

The nonparametric equation of state representation based on a Gaussian process conditioned on a set of 
nuclear models~\cite{Landry:2018prl,Essick:2019ldf}, including models with strong phase transitions, 
yields $R_{1.4}=10.95^{+2.00}_{-1.37}$km. This measurement is based on a hierarchical analysis of the mass measurements of the heavy 
pulsars and both gravitational wave signals GW170817 and GW190425~\cite{Landry:2020vaw}, though GW190425 is found to be uninformative
as also shown in~\cite{Abbott:2020uma}. Figure~\ref{fig:radius} shows the 90\% credible intervals on the mass-radius and the tidal
deformability-mass planes for the analysis prior (black), an analysis of only the heavy pulsars (cyan), an analysis of only the two gravitational wave
 signals (green), and the combined constraints (purple). The analysis prior is very wide, allowing for radii as low as $8$km and as high as $15$km
 for $m=1.4M_{\odot}$, so it reveals the influence of each data set. The heavy pulsars serve to rule out the soft part of the equation of state
 and the lower radii while gravitational waves -primarily GW170817- rule out stiff equations
 of state with large radii. Direct model selection between nuclear models also confirms this, as stiff models are disfavored~\cite{LIGOScientific:2019eut}.

Equation of state representations that are anchored to low- or extremely high-density nuclear calculations can lead to even more stringent constraints,
perhaps at the expense of model-dependency and restricted applicability to within a specific nuclear framework. On the low density side, the equation of state is commonly 
represented with results from chiral effective field theory 
calculations~\cite{Tews:2012fj,Drischler:2016djf,Drischler:2017wtt,Tews:2018kmu,Drischler:2020yad} up to a certain
 density that is comparable to the nuclear saturation density. 
On the opposite side, perturbative quantum chromodynamics calculations~\cite{Kurkela:2009gj} describe the limiting equation of state
behavior for extremely high densities (higher than central neutron star densities). The in-between region can be described 
with a generic equation of state representation, see Fig. 3 of~\cite{Annala:2017llu}, while the transition density between chiral 
effective field theory and the generic representation
dictates the extent to which the resulting analysis depends on the nuclear calculations.

Unsurprisingly, the higher the density to which chiral effective field theory calculations are used,
the more stringent the resulting constraints are. Employing the speed of sound parametrization and gravitational wave data from GW170817, while
imposing a sharp cutoff on the maximum mass of $1.9M_{\odot}$, Ref.~\cite{Capano:2019eae} finds that a transition density at twice compared to 
once the nuclear saturation reduces the radius uncertainty by $\sim500$m on the high side.
The dependence on the choice of density up to which chiral effective field theory calculations are used can be marginalized over based
on the data.
Using the nonparametric equation of state representation and marginalizing over the transition density between half and twice saturation
leads to $R_{1.4}=11.40^{+1.38}_{-1.04}$km at the 90\% level when using information about the maximum neutron star mass
and gravitational waves~\cite{Essick:2020flb}. Comparing to~\cite{Landry:2020vaw} suggests that use of chiral effective field theory restricts
the fiducial radius by about $1$km on the low side. 

Further studies have used similar equation of state representations combining chiral effective field theory and generic representations, while
incorporating information from the gravitational wave data in terms of a sharp constraint at the 90\% credible upper limit of $\tilde{\Lambda}$,
rather than the full four-dimensional posterior over masses and tides. For example, Ref.~\cite{Annala:2017llu} combined the tidal deformability
bound $\tilde{\Lambda}<800$ and a maximum mass constraint of $2M_{\odot}$ with an equation of state model in terms of chiral effective theory,
piecewise polytropes, and perturbative quantum chromodynamics calculations at large densities
to conclude that $R_{1.4}<13.6$km and $\Lambda_{1.4}>120$. Similar calculations employing both piecewise polytropes and the speed of
sound parametrization to capture potential strong phase transitions lead to 
$12.00<R_{1.4}/\mathrm{km}<13.45$ and $8.53<R_{1.4}/\mathrm{km}<13.74$
respectively~\cite{Most:2018hfd}. Finally, Refs.~\cite{Tews:2018chv,Tews:2019cap} use the speed of sound parametrization and chiral effective
field theory up to once and twice saturation to find $8.4 < R_{1.4}/\mathrm{km} < 15.2$  and $8.7 < R_{1.4}/\mathrm{km} < 12.6$ respectively.

Finally, relativistic-mean-field models have been used to compare the experimentally probed properties of finite 
nuclei to the recent gravitational wave data~\cite{Fattoyev:2017jql,Nandi:2018ami,Malik:2018zcf,Sabatucci:2020xwt,Posfay:2020xgp,Traversi:2020aaa}.
The main terrestrial constraint is the neutron skin thickness $R_{\mathrm{s}}$ 
of $^{208}$Pb~\cite{Horowitz:2000xj,Vidana:2009is,Abrahamyan:2012gp,Horowitz:2012tj}, the difference between the root-mean-square 
radii for neutrons and protons in the $^{208}$Pb nucleus. Comparing relativistic mean field models to the 
$\Lambda_{1.4}<800$~\cite{TheLIGOScientific:2017qsa} constraint from GW170817 yields $R_{1.4}<13.8$km~\cite{Fattoyev:2017jql},
while the updated estimate $\Lambda_{1.4}<580$~\cite{Abbott:2018exr} yields $R_{1.4}<12.9$km~\cite{Nandi:2018ami}.
The updated value leads to $R_{\mathrm{s}}\lesssim0.20$fm~\cite{Nandi:2018ami}, consistent but on the low side of
 the $R_{\mathrm{s}}=0.33^{+0.16}_{-0.18}$fm measurement from PREX~\cite{Abrahamyan:2012gp}. 
 A potential tightening of the low limit on $R_{\mathrm{s}}$
 by PREXII could signal a sudden softening of the equation of state at high densities and hint towards a phase transition.

Figure~\ref{fig:radius} shows the radius constraints discussed above (and some discussed later) with horizontal lines of arbitrary ordinate. 

%%%%%%%%%%%%-------------------------------------------------
\subsubsection{Microscopic properties}

Besides radius/tidal constraints that take into account the possibility of a strong phase transition in the equation of state
and multiple branches on the mass-radius plane (solid lines in Fig.~\ref{fig:radius}), astronomical observations can also be used to place
constraints on the relevant parameter space.
In general, since GW170817 provided an upper limit on $\tilde{\Lambda}$, any physical effect that results in a softening of the equation of state
can be consistent with the 
data~\cite{Souza:2020eyq,Das:2018frc,Quddus:2019ghy,Dexheimer:2018dhb,Li:2019tjx,Ribes:2019kno,Sun:2018tmw,Li:2019sxd}
As such, a strong first-order phase transition might make an equation of state model compatible with the GW170817 data, even if the hadronic
part on its own is not~\cite{Paschalidis:2017qmb,Nandi:2017rhy,Han:2018mtj,Han:2019bub,Ferreira:2020evu}. A number of studies have constructed
models that can successfully interpret the inspiral signal from GW170817 as the coalescence of any combination of hadronic and hybrid 
hadronic-quark neutron stars and place corresponding constraints on the relevant model parameter
 space~\cite{Paschalidis:2017qmb,Nandi:2017rhy,Zhou:2017pha,Christian:2018jyd,Li:2018ayl,Alvarez-Castillo:2018pve,Gomes:2018eiv,Montana:2018bkb,Sen:2018yyq,Sieniawska:2018zzj,Ayriyan:2018blj,Christian:2019qer,Blaschke:2020qqj,Miao:2020yjk}.

Establishing the presence of a strong phase transition without resorting to a specific nuclear model for either the hadronic or the quark part
of the equation of state is hindered by the small number of observations available. Analyses relying on nonparametric equation of state
representations are generic enough to account for phase transitions, but the final result is expressed in terms of pressure-density or mass-radius
curves, which requires further interpretation. One possible interpretation 
consists of counting the number of stable branches an equation of state possesses
on the mass-radius plane~\cite{Essick:2019ldf}, where the presence of multiple branches is sufficient (but not necessary) condition for a
strong phase transition. Data from heavy pulsars, gravitational waves, and NICER result in a very weak preference for such multiple branches
at the level of $2:1$~\cite{Landry:2020vaw}. In the case of a strong phase transition, the equation of state is consistent with a softening around
saturation density, followed by a stiffening around twice saturation, see Fig. 3 in~\cite{Landry:2020vaw}. 

Another indication about microscopic interactions comes from the speed of sound $c_s$ inside neutron stars. 
The combination of a large maximum
mass and a soft low-density equation of state hints towards a violation of the conformal, weakly-interacting 
limit of $c_s^2<1/3$~\cite{PhysRevLett.114.031103,Tews_2018SOUNDSPEED}. Since GW170817 is also consistent with a soft low-density 
equation of state, it serves to strengthen this conclusion~\cite{Reed:2019ezm,Landry:2020vaw}. However, though a violation of the conformal
limit is an indication of strongly-interacting matter, this is not equivalent to the presence of quark cores in neutron stars~\cite{Annala:2019puf}.

Finally, astronomical observations and terrestrial experiments can be translated to microscopic 
constraints through the symmetry energy~\cite{Li:2019xxz},
the energy budget for increasing asymmetry between neutrons and protons. Various studies have empirically identified correlations
between various combinations of parameters that characterize the symmetry energy
and its density dependence with the neutron star radius and tidal parameters~\cite{Fattoyev:2013rga,Alam:2016cli,Zhang:2018vbw,Li:2019sxd,Guven:2020dok}
or terrestrial experimental results~\cite{Vidana:2009is,Tamii:2011pv,Tsang:2008fd,Tsang:2012se,Agrawal:2020wqj}, as they all are linked to the low-density behavior of the equation of state around (twice) saturation. The tidal constraints from GW170817,
and more recently radius constraints from NICER have been used to examine implications for the symmetry 
energy~\cite{Zhang:2018vrx,Carson:2018xri,Tsang:2019vxn,Tsang:2019mlz,Xie:2019sqb,Raithel:2019ejc,Ferreira:2019bgy,Zimmerman:2020eho,Zhang:2020azr,Xie:2020tdo,Routray:2020zkf}
as well as potential systematics in the mapping~\cite{Carson:2019xxz}.

%%%%%%%%%%%%-------------------------------------------------
\subsection{Multimessenger constraints}

The properties of the electromagnetic counterpart to GW170817 depend on the outcome of the coalescence, which in turn depends primarily
on the binary mass (which can be measured from the inspiral signal) and the equation of state. Any subsequent constraints about the
equation of state from the electromagnetic signal will inevitably depend on the merger and postmerger modeling of the system, the mechanism
that gives rise to the electromagnetic emission, as well as its interpretation. Though less understood than the gravitational wave emission, 
progress has been made on the above. The result is a lower limit on $\tilde{\Lambda}$ and the neutron star radius, which 
-if robust against systematic errors- is nicely complementary to the inspiral constraints from the gravitational wave signal
as those provide an upper limit on the neutron star compactness. 

Energetic arguments suggest that the remnant star formed after the merger was initially a hypermassive neutron star which eventually collapsed into
a black hole~\cite{Margalit:2017dij}, though alternative interpretations and their implications have been explored~\cite{Mosta:2020hlh,Ai:2018jtv,Ai:2019rre,Wu:2020zhr}. 
A number of studies have used this and different assumptions about the post-merger evolution of the system
to suggest an upper limit on the maximum mass of stable, nonrotating neutron stars of $\lesssim 2.3 M_{\odot}$~\cite{Margalit:2017dij,2018ApJ...852L..25R,Ruiz:2017due,Shibata:2017xdx,Shibata:2019ctb,LIGOScientific:2019eut,Ai:2019rre,Shao:2019ioq}.
Using a similar interpretation, a relation between the radius, the maximum neutron star mass, and the threshold mass for prompt collapse
of the merger remnant~\cite{Bauswein:2013jpa} suggests $R_{1.6}\gtrsim 10.6$km~\cite{Bauswein:2017vtn}. The arguments leading to the 
above constraints rely on equations of state without strong phase transitions and the possibility of 
hybrid quark-hadron stars~\cite{Drago:2018nzf,Bozzola:2019tit,Bauswein:2020aag},
however, detection of further systems can provide stronger constraints or put the above expectations to the test~\cite{Margalit:2019dpi}.

The observed kilonova after GW170817 was powered by material ejected from the merger~\cite{Metzger:2019zeh}, 
the properties of which depend on the
neutron star tidal deformability. A small value of $\tilde{\Lambda}$ corresponds to more compact
stars whose merger results in a remnant that promptly collapses to a black hole. This prompt collapse inhibits the mass ejection that 
is needed in order
to power the observed kilonova. Hence, the electromagnetic counterpart in this case provides a lower limit on the compactness of the 
merging neutron stars, in line with~\cite{Bauswein:2017vtn}. 
Results from numerical simulations have been used to quantify the above through a relation between the
disk mass~\cite{Radice:2017lry,Radice:2018ozg} or ejecta mass and velocity~\cite{Coughlin:2018miv,Coughlin:2018fis} 
with $\tilde{\Lambda}$.
The resulting constraint amounts to $\tilde{\Lambda}\gtrsim300$, though the 
quantitative accuracy of the fitting formulas has been the subject of 
debate~\cite{Radice:2018ozg,Coughlin:2018fis,Kiuchi:2019lls,Bauswein:2020aag}. 
A similar lower limit on $\tilde{\Lambda}$ has also been proposed on the basis of the observed gamma ray burst~\cite{Wang:2018nye}.
Both~\cite{Radice:2018ozg} and~\cite{Coughlin:2018fis} achieve comparable radius constraints 
$11.1\lesssim R_{1.4}/\mathrm{km}\lesssim 13.7$, 
also shown in Fig.~\ref{fig:radius}, where the lower (upper) limit is informed by the electromagnetic (gravitational) signal.
Followup studies using 
relativistic mean field theory models~\cite{Malik:2018zcf} and the relation between $\tilde{\Lambda}$ and radius~\cite{Burgio:2018yix} 
arrive at similar conclusions.
As with the maximum mass constraints above, this result
is also only valid for hadronic equations of state, as numerical simulations of merging hybrid stars suggest that the remnant collapse behavior
and mass ejection is different~\cite{DePietri:2019khb}, while magnetic fields could either stabilize or destabilize a hybrid 
star~\cite{Gomes:2018bpw}.

%%%%%%%%%%%%-------------------------------------------------
\subsection{Incorporating constraints from NICER and X-ray binaries}

We conclude the discussion of observational constraints on the equation of state using GW170817 with a brief discussion of 
results that incorporate mass-radius measurements from accreting low mass X-ray binaries~\cite{Ozel:2016oaf} and isolated neutron stars
with NICER~\cite{Miller:2019cac,Riley:2019yda}. On the X-ray binary front, Ref.~\cite{Kumar:2019xgp} used equation of state insensitive
relations~\cite{Yagi:2016bkt} to translate between radii constraints from X-ray binaries and tidal constraints from GW170817 and arrived
at $\Lambda_{1.4}=196^{+92}_{-63}$ at the 90\% level. A similar improvement in the overall constraints was reported in~\cite{Fasano:2019zwm},
who use the spectral equation of state representation to find $R_1 =12.4^{+0.5}_{-0.4}$km for the more massive GW170817 binary component,
and similar for the companion. The above are a considerable improvement compared to GW170817-only results, but at the potential
expense of additional systematic errors. 

Finally, the recent mass-radius constraint from J0030+0451, an isolated pulsar, by NICER~\cite{Miller:2019cac,Riley:2019yda}
has led to efforts to combine constraints not only with gravitational waves, but also nuclear data and 
X-ray binaries~\cite{Jiang:2019eiq,Jiang:2019rcw}, as well as the GW170817 counterpart~\cite{Dietrich:2020lps}. The overall
picture suggests that results from GW170817 and J0030+0451 are consistent with each other, though the former (latter) hint towards slightly
softer(stiffer) equations of state~\cite{Miller:2019cac,Raaijmakers:2019qny,Raaijmakers:2019dks,Landry:2020vaw}. 
Using the nonparametric equation of state representation with data from heavy pulsars, GW170817, GW190425,
and J0030+0451, Ref.~\cite{Landry:2020vaw} finds $R_{1.4} = 12.32^{+1.09}_{-1.47}$km, also shown in Fig.~\ref{fig:radius}. 
A similar analysis was presented 
in~\cite{Miller:2019cac} by means of the spectral and piecewise polytropic representations, though without quantitative results.
The piecewise polytropic and speed-of-sound representations were instead employed in~\cite{Raaijmakers:2019qny,Raaijmakers:2019dks},
though the chosen priors on the equation of state were so narrow that they overwhelmed the resulting 
constraints, prohibiting direct comparison with~\cite{Landry:2020vaw}.

The above examples show how combining information from gravitational waves, NICER, and potentially a future moment
of inertia measurement~\cite{Lattimer_2005,Landry:2018jyg,Miller:2019nzo,Landry:2020vaw,Greif:2020pju} can strengthen overall constraints.
At the same time, it is worth remembering that neutron stars are extreme relativistic objects with which the full theory of general
relativity can be tested. A potential disagreement between the different messengers, beyond what can be accounted for by nuclear physics 
uncertainties could signal a breakdown of the theory. For the case of GW170817 and J0030+0451 it has been shown that their measured properties
agree with each other when recast in terms of equation of state insensitive relations between the compactness, the moment of inertial
and the tidal deformability~\cite{Silva:2020acr}, placing constraints on potential deviations from general relativity.

%%%%%%%%%%%%%%%%%%%%%%%%%%%%%%%
\section{Future challenges and opportunities}
\label{future}

The detection of GW170817 introduced the tidal properties of neutron stars as a complementary probe of their structure to the more
traditional observables of masses and radii. Progress in the intervening years has been rapid, establishing gravitational waves as a 
promising input for nuclear models, which are now routinely tuned, compared, and constrained against GW170817,
e.g.~\cite{Annala:2017tqz,Zhu:2018ona,Sun:2018tmw,Lim:2018bkq,Lai:2018ugk,Krastev:2018wtx,Tews:2019qhd,Marczenko:2020jma,Zhao:2020dvu,Wang:2020dov,Traversi:2020aaa,Alvarez-Castillo:2020fyn,Adam:2020yfv}.
The continued observational campaigns and sensitivity improvements~\cite{Aasi:2013wya} as well as the addition of new
planned or proposed detectors~\cite{Aso:2013eba,LIGOINDIA,2010CQGra..27h4007P,2011CQGra..28i4013H,ISwhitePaper}
in the global gravitational wave detector network both opens up unique opportunities and brings novel challenges to the surface.

The main obstacle towards an anticipated ${\cal{O}}(1)$km radius constraint in the coming years~\cite{Landry:2020vaw} is that of
systematic biases in the analyses, the dominant of which is related to waveform model inaccuracies~\cite{Wade:2014vqa}.
Other sources of systematic uncertainties in gravitational wave analyses are the detector calibration~\cite{Sun:2020wke} and noise 
modeling~\cite{Chatziioannou:2019zvs}, however the former affects the waveform phase to within a few degrees 
(compare this to Fig.~\ref{fig:waveform} showing how tides affect the waveform phase by multiple cycles) and the latter mostly affects the 
inferred amplitude of the signal. Despite the signal strength, analyses with different waveform 
models~\cite{Dietrich:2017aum,Dietrich:2018uni,Kawaguchi:2018gvj,Kiuchi:2017pte,Kiuchi:2019kzt,Nagar:2018zoe,Hinderer:2016eia,Steinhoff:2016rfi} 
suggest that waveform systematics are subdominant to statistical errors for 
GW170817~\cite{Abbott:2018wiz,2018arXiv181112907T,Narikawa:2019xng} and similar for GW190425~\cite{Abbott:2020uma}. 
However, a potential future observation with a signal-to-noise
ratio  $\sim100$ (which is comparable the case of GW170817 observed with detectors that have reached their design sensitivity)
would result to tidal inference that is dominated by systematic waveform uncertainties~\cite{Dudi:2018jzn,Samajdar:2018dcx}.

At the same time, more and louder detections offer the opportunity of going beyond a single radius measurement~\cite{Landry:2020vaw} or even
a characterization of features of the equation of state such as phase transitions~\cite{Chatziioannou:2019yko}. Improved detector sensitivity
will allow us to measure higher order terms beyond the $\ell=2$ tidal deformability~\cite{Jimenez-Forteza:2018buh},
directly observe the postmerger signal~\cite{Torres-Rivas:2018svp}, and study the properties of the neutron star crust in terms
of its elastic properties~\cite{Pereira:2020jgv,Gittins:2020mll}, heating~\cite{Pan:2020tht}, and shattering~\cite{2012PhRvL.108a1102T}.
Another effect of interest is the potential resonant excitation of different neutron star modes driven by the
 orbital motion~\cite{Flanagan:2006sb,Poisson:2020eki}.
In spite of its high resonance frequency, ${\cal{O}}(1000)$Hz, an f-mode excitation could be detected in the
 future~\cite{Schmidt:2019wrl,Pratten:2019sed,Wen:2019ouw}, potentially aided by orbital eccentricity~\cite{Chirenti:2016xys} 
or neutron star spins~\cite{Ma:2020rak}. A search for a proposed nonresonant mode
 coupling~\cite{Venumadhav2013,Weinberg2016,Zhou2017,Essick:2016tkn} in GW170817 and GW190425 was
 inconclusive~\cite{Weinberg:2018icl,Abbott:2020uma}.

As a concluding remark, combining information from multiple messengers
offers the clearest pathway towards stringent equation of state constraints and probes of neutron star
structure. This allows us not only  to decrease statistical errors, but also to assess the impact of systematics on the modeling
of very diverse astrophysical phenomena.

%%%%%%%%%%%%%%%%%%%%%%%%%%%%%%%
\section{Acknowledgements}
\label{acks}

We thank Sophia Han for useful comments on this manuscript, as well as 
the LIGO and Virgo Collaboration Extreme Matter working
group for invaluable discussions over the years.  The Flatiron Institute is supported by
the Simons Foundation.

\bibliographystyle{spphys}
\bibliography{refs}

\end{document}